\documentclass[twocolumn]{aastex7}

\usepackage[colorinlistoftodos]{todonotes}
\usepackage{amssymb}
\usepackage{comment}
\usepackage{multirow,acronym}
\usepackage{natbib}
\usepackage{makecell}
\usepackage{booktabs}
\usepackage{amsmath}
\usepackage{subfigure}
\usepackage{color}
\usepackage{xcolor}
\usepackage{hyperref}
\usepackage[T1]{fontenc}
\usepackage{graphicx}
\usepackage{bm}
\usepackage{CJK}

\hypersetup{colorlinks=true, citecolor=blue, 
linkcolor=cyan, urlcolor=magenta}

\usepackage{float}
\usepackage{amsmath,amssymb}
\usepackage{hyperref} 
\usepackage{graphicx} 
\usepackage{color}
\usepackage{verbatim}    
\usepackage{enumitem}
\usepackage{natbib}
\usepackage{multirow}

\newcommand{\proptosim}{\mathrel{\vcenter{
 \offinterlineskip\halign{\hfil$##$\cr
 \propto\cr\noalign{\kern2pt}\sim\cr\noalign{\kern-2pt}}}}}
\renewcommand{\b}[1]{\boldsymbol{#1}}

\newcommand{\response}[1]{{\bf\color{blue} #1}}

\renewcommand{\min}{\mathrm{min}}
\renewcommand{\max}{\mathrm{max}}

\newcommand{\m}{\mathrm{m}}

\newcommand{\g}{\mathrm{g}}
\newcommand{\G}{\mathrm{G}}
\newcommand{\K}{\mathrm{K}}

\newcommand{\s}{\mathrm{s}}

\renewcommand{\d}{\mathrm{d}}
\renewcommand{\m}{\mathrm{m}}
\newcommand{\f}{\mathrm{f}}
\newcommand{\e}{\mathrm{e}}

\renewcommand{\roman}[1]{\textsc{\MakeLowercase{#1}}}

\newcommand{\p}{{\rm p}}

\renewcommand{\response}[1]{#1} 
\newcommand{\DOA} {Department of Astronomy, School of
  Physics, Peking University, Beijing 100871, China}
\newcommand{\KIAA}{Kavli Institute for Astronomy and
  Astrophysics, Peking University, Beijing 100871, China}

\newcommand{\w}{{\rm w}}

\acrodef{OC}{open cluster}
\acrodef{BH}{black hole}
\acrodef{DF}{dynamic friction}
\acrodef{NDF}{negative dynamic friction}
\acrodef{IMF}{initial mass function}
\acrodef{AGN}{active galactic nucleus}

\begin{document}

\title{ AGN star dynamics under the Influence of
  Outflow-Ambient Interactions}

\author[0009-0003-9343-8107]{Muxin Liu}
\affiliation{\DOA}
\affiliation{\KIAA}
\email{liumxwsh@stu.pku.edu.cn}

\author[0000-0002-6540-7042]{Lile Wang}
\affiliation{\KIAA}
\affiliation{\DOA}
\email{lilew@pku.edu.cn}

\author{Peng Peng}
\affiliation{\DOA}
\affiliation{\KIAA}
\email{peng.p@pku.edu.cn}

\correspondingauthor{Lile Wang} 
\email{lilew@pku.edu.cn}

\begin{abstract}
  Stars with outflows interacting with ambient gas
  experience accelerations arising from the gravitational
  feedback induced by the interaction structure. In this
  work, three-dimensional (3D) local shearing box
  simulations are performed to investigate the dynamical
  evolution of a star with outflows embedded in the outer
  regions of an active galactic nucleus (AGN) disk. Two
  types of stellar wind are considered: isotropic winds and
  axisymmetric jets, along with variations in the radial
  pressure gradient profile. The results show that
  anti-friction enables AGN stars to acquire angular
  momentum from the ambient gas, resulting in outward
  migration away from the disk center. The formation and
  stability of the head-wind structure, which is crucial for
  maintaining anti-friction, are sensitive to both the
  strength of the stellar outflow and the radial pressure
  gradient of the disk gas. Once the head-wind structure is
  disrupted, the anti-friction effect ceases to operate
  effectively. A case study is also presented, focusing on a
  stellar-mass black hole (sBH) in an AGN disk. It is shown
  that jet material launched along the z-axis is confined to
  the trailing side of the object's motion by high gas
  inflow velocities, thereby activating anti-friction and
  inducing outward migration. If such an sBH migrates inward
  initially, the interplay between inward and outward
  migration may trap it at an equilibrium radius,
  potentially facilitating the formation and merger of black
  hole binaries.
\end{abstract}


\section{Introduction}
\label{sec:intro}
The discovery of quasars marked a significant breakthrough
in understanding the formation and evolution of galaxies
\citep{1963Natur.197.1040S}, revealing them as the most
luminous members of active galactic nuclei (AGNs;
\citealt{1997iagn.book.....P,2008ARA&A..46..475H}). These AGNs are believed to
be powered by the rapid accretion of matter onto
supermassive black holes (SMBHs), where gravitational energy
is efficiently released through accretion disks at high
rates \citep{1969Natur.223..690L}. Given their crucial role
in AGNs, the structure and dynamics of these disks have been
extensively studied since the pioneering work of
\citet{1973A&A....24..337S}.

Within AGN disks, the mechanisms governing the presence and
motion of stars and compact stellar remnants have long been
a focus of theoretical research. These objects enter AGN
disks via at least two mechanisms. One mechanism involves
{\it in-situ} formation: when an extended AGN accretion disk
becomes self-gravitating, it may undergo fragmentation,
leading to the birth of stars \response{\citep{1978AcA....28...91P,
  1980SvAL....6..357K, 1987Natur.329..810S,
  2003MNRAS.339..937G,2011ApJ...730...45J,2023ApJ...948..120C}}. This process preferentially produces
massive stars, which can grow further through sustained
accretion \citep{2003astro.ph..7084L, 2004ApJ...608..108G,
  2020MNRAS.493.3732D}.  The other mechanism is the stellar
capture, during which stars originally orbiting near the
galactic center may lose energy and angular momentum through
repeated interactions with the AGN disk. Such dissipative
processes, driven by hydrodynamic drag and the excitation of
resonant density and bending waves, settle stars into
circular, corotating orbits within the disk
\response{\citep{1991MNRAS.250..505S,1993ApJ...409..592A,2020ApJ...889...94M,2024MNRAS.528.4958W}}. This
capture process is expected to be efficient within
approximately 10 pc of the SMBH \citep{1993ApJ...409..592A,
  2012ApJ...758...51J, 2016MNRAS.460..240K,
  2020MNRAS.499.2608F, 2020ApJ...889...94M}. Depending on
the AGN's lifetime and disk properties, simulations suggest
that a few hundred stars could be captured over a period of
about one million years
\citep{2018MNRAS.476.4224P,2020MNRAS.499.2608F}. Over longer
timescales, approaching 100 million years, the number of
captured stars could exceed $10^4$
\citep{1993ApJ...409..592A}.

Stars embedded in AGN disks (AGN stars) are expected to
travel under conditions quite different from those in
standard galactic environments. Due to the extreme
temperatures and densities of AGN disks, these stars are
subject to very different boundary conditions (BCs) than
normal stars, which significantly influence their dynamical
evolution. Consequently, the detailed dynamical evolution of
AGN stars remains an open question. Proper modeling must
account for their prolonged exposure to intense pressures,
temperatures, and accretion rates. Several studies have
explored various aspects of the dynamical behavior of AGN
stars. \citet{2021ApJ...910...94C} introduced modifications
to standard stellar dynamical evolution models to
accommodate the extreme conditions present in different
regions of AGN disks. Additionally, the dynamics of
stellar-mass binary black holes in AGN disks have been
examined by \citet{2022MNRAS.517.1602L, 2023MNRAS.522.1881L,
  2024MNRAS.529..348L}, while \citet{2023ApJ...950....3P,
  2024arXiv241116070P} investigated the interaction between
a gap-opening intermediate-mass-ratio inspiral (IMRI) and
surrounding sBHs within an AGN disk. The high-density
environment of AGN disks facilitates conditions where stars
may exceed the Eddington limit, leading to the production of
powerful outflows \citep{2021ApJ...911L..14W}. However,
existing studies have not yet addressed how stellar outflows
influence the dynamical evolution of AGN stars. This work
aims to incorporate the effects of outflows into the study
of AGN star dynamics.

It is well acknowledged that a massive object moving through
a gaseous medium experiences dynamical friction
\citep{1942ApJ....95..489C, 1999ApJ...513..252O,
  2004NewAR..48..843E}, a key hydrodynamic drag mechanism
involved in stellar capture. However, this scenario changes
if the object launches strong
winds. \citet{2020MNRAS.492.2755G} demonstrate that when the
wind velocity is sufficiently high, a large underdense
region forms around the object, altering the net
gravitational force exerted by the ambient gas. This effect
can reverse the drag force, causing it to align with the
object's motion—a phenomenon known as anti-friction, where
the object accelerates instead of
decelerating. \citet{2020MNRAS.494.2327L} employed
hydrodynamic simulations to investigate how stellar outflows
affect accretion rates and modify the strength of
anti-friction. Moreover, \citet{2022ApJ...932..108W} studied
anti-friction in binary systems using global 3D hydrodynamic
simulations. Liu et al. further examined the role of
anti-friction in the dynamical evolution of stars in open
clusters. \response{\citet{2025MNRAS.540.2952C} investigate the impact of dynamical friction on stellar motion using a series of two-dimensional simulations in a wind tunnel configuration.} This study performs three-dimensional (3D) local
simulations of a star with outflows embedded in an AGN disk
using the shearing box approximation, which was initially
introduced by \citet{1995ApJ...440..742H} in the study of
the magnetorotational instability, re-casting the equations
of motion in a local Cartesian frame co-rotating with the
disk at an arbitrary radius $r_0$. The validity of this
approximation is maintained as long as the studied domain
remains significantly smaller than $r_0$, making it a
valuable tool for investigating local accretion disk
dynamics \citep{2003LNP...614..329B}.

This paper is structured as follows. \S \ref{sec:method}
outlines the numerical methods, disk model, and all
simulated cases. \S \ref{sec:result-fiducial} compares the
effects of friction and anti-friction, analyzing the
physical parameters that influence gas-star interactions. \S
\ref{sec:black-hole} presents a case study on the dynamical
evolution of an sBH embedded in an AGN disk under the
influence of anti-friction. Finally, \S \ref{sec:summary}
provides a discussion and summary of the results.

\begin{deluxetable}{lr}
  \tablecolumns{2}
  \tabletypesize{\footnotesize}
  \tablewidth{0pt}
  \setlength{\tabcolsep}{10pt}
  \tablecaption{Initial Properties of the Fiducial Model}
  \label{tab:fiducial-property}
  \tablehead{ \multicolumn{1}{l}{Item} &
    \multicolumn{1}{r}{Value} } \startdata
  \textbf{Disk Properties} & \\
  $M$ & $10^8~M_\odot$ \\
  $\rho_0$ & $\sim8\times10^{-12}~{\rm g~cm^{-3}}$ \\
  $c_{\rm iso}$ & \response{$10^6~{\rm cm~s^{-1}}$} \\
  $x_{\rm p}$ & $90~{\rm AU}$ \\
  $v_{\rm acc}^*$ & $0~{\rm cm~s^{-1}}$ \\
  \textbf{Stellar Properties} & \\
  $m$ & $8~M_\odot$ \\
  $r_0$ & $5000~R_{\rm sch}$ \\
  $\dot{m}$ & $\sim3\times10^{-3}~M_\odot~{\rm yr^{-1}}$ \\
  $T_{\rm outflow}$ & $\sim6\times10^4~{\rm K}$ \\
  $v_{\rm src}$ & $\sim8\times10^7~{\rm cm~s^{-1}}$ \\
  $r_{\rm soft}/r_{\rm src}$ & $\sim3$ \\
  Outflow Type$^\dagger$ & Isotropic \\
  \textbf{Simulation Parameters} & \\
  $L_x, L_y, L_z$ & $75, 150, 75~{\rm AU}$ \\
  Resolution $(N_x, N_y, N_z)$ & $128, 256, 128$ \\
  $t_{\rm evo}$ & $15~P$ \\
  \enddata \tablecomments{
    $^*$: $v_{\rm acc} = 0$ corresponds to a non-accreting disk.\\
    $^\dagger$: The fiducial model assumes an isotropic
    stellar outflow. Jet and no-outflow models are also
    discussed (see \S\ref{sec:jet-effect}).  }
\end{deluxetable}

\begin{deluxetable*}{ccccc}
  \tablecolumns{2} \tabletypesize{\scriptsize}
  \tablewidth{500pt}
  \tablecaption{ ALL simulated models in this work.}
  \label{tab:model-summary}  
  \tablehead{ \colhead{Series} & \colhead{Name} &
    \colhead{$^{**}\langle{a_y}\rangle~[10^{-8}~{\rm
        km}~{\rm s}^{-2}]$}&
    \colhead{$^{**}\langle\dot{r}_{\rm cir}\rangle~[{\rm
        km}~{\rm s^{-1}}]$}&\colhead{Description} }
  \startdata & \response{FID$^{+-}$} &0.24&2.41
  &\response{Fiducial model; $x_{\rm p} = 90~{\rm AU}$ (\S
    \ref{sec:gas-star-interaction})}
  \\
  & \response{MASS$^{+}$} &\response{-0.30} &\response{-2.96} & \response{$m = 80\,M_\odot,\,v_{\rm src}=2\times10^8~{\rm cm~s^{-1}}$(\S \ref{sec:gas-star-interaction})} \\
  & \response{GAMMA$^{+}$} &\response{0.08} &\response{0.79} &\response{ $\gamma = 4/3$(\S \ref{sec:gamma-section})} \\
  & \response{PG-45$^{+-}$} &0.77 &7.59 & $x_{\rm p} = 45~{\rm AU}$(\S \ref{sec:pressure-gradient-effect}) \\
  Fiducial& \response{PG-0$^{+-}$}  &-0.37 &-3.65 & $x_{\rm p} = 0~{\rm AU}$(\S \ref{sec:pressure-gradient-effect}) \\
  Studies$^\dagger$ & \response{JET-x$^{+}$}  &-0.14 &-1.38 & Jet propagation along $\pm x$ (\S \ref{sec:jet-effect})\\
  & \response{JET-y$^{+}$}  &-0.26 &-2.56 & Jet propagation along $\pm y$ (\S \ref{sec:jet-effect})\\
  & \response{JET-z$^{+}$}  &-0.24 &-2.36 & Jet propagation along $\pm z$ (\S \ref{sec:jet-effect})\\
  & \response{ACCD$^{+-}$}  &0.11 &1.1 & Considering accretion disk (\S \ref{sec:accretion-disk})\\
  \hline
  Case Study$^*$& \response{BH-jet-z$^{+}$}  &0.40 &1.83 & Considering a jet along $\pm z$(\S \ref{sec:black-hole})\\
  \enddata \tablecomments{ $^\dagger$:
    $t_{\rm evo} = 15~{P}$.  $^*$: $t_{\rm evo} =
    0.1~{P}$. $^{**}$: The time-averaged azimuthal
    acceleration $\langle{a_y}\rangle$ and the mean rate of
    change of the orbital radius
    $\langle\dot{r}_{\rm cir}\rangle$ under the influence of
    anti-friction.\response{$^+$: Anti-friction cases where
      stars exhibit outflows (model names contain
      ``-anti''). $^-$: Friction cases where stars lack
      outflows (model names contain ``-fric'').} }
\end{deluxetable*}

\begin{figure*} 
  \centering    
  \includegraphics[width=0.95\columnwidth]
{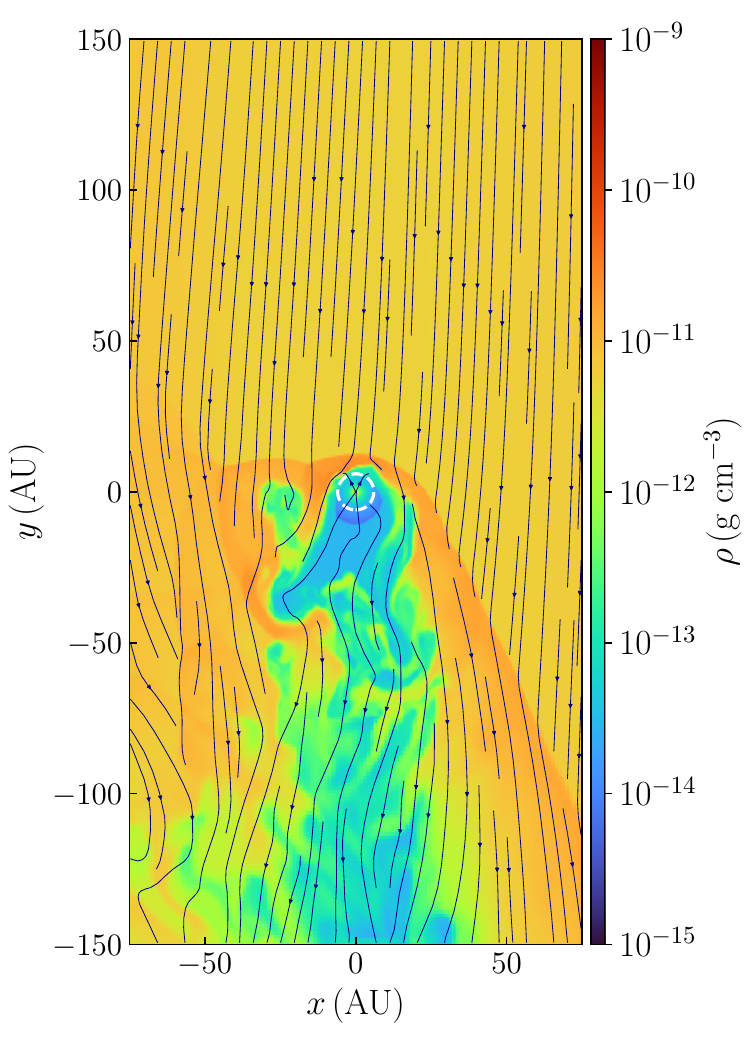}
  \includegraphics[width=0.95\columnwidth]
{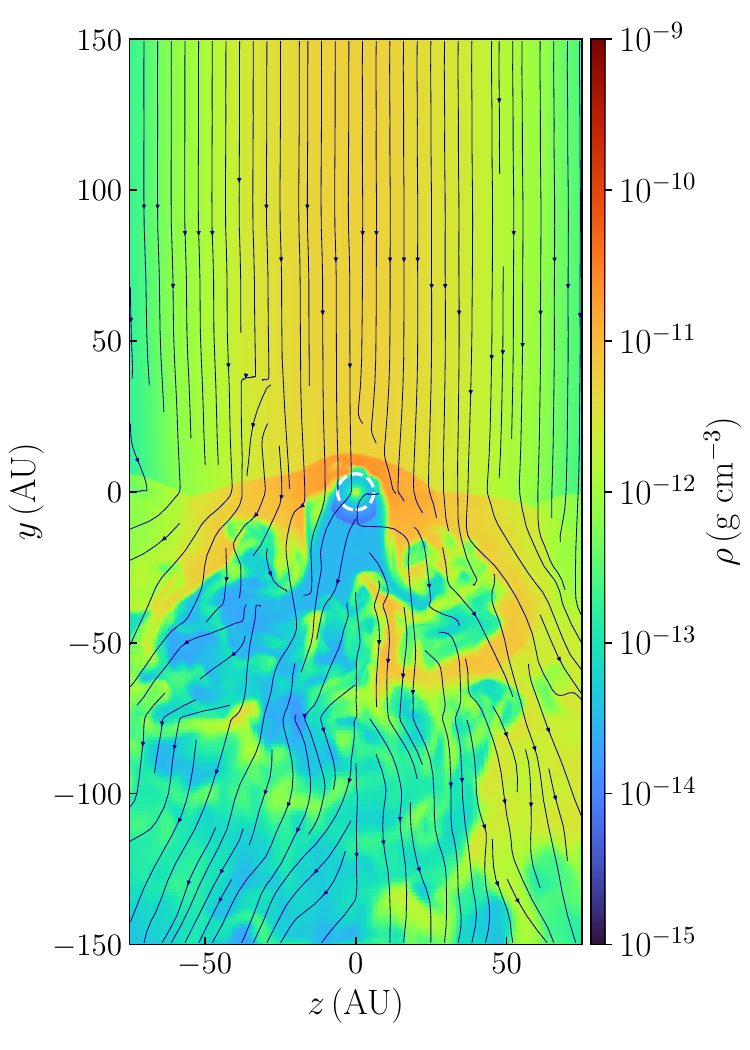}
\caption{Gas density $\rho$ in the $z=0$ plane (i.e., the
  orbital plane; left column) and the $x=0$ plane (right
  column) at the $t_{\rm evo} = 15~P$ snapshot of the
  fiducial model. The star is located at the origin, with
  color indicating gas density and black velocity
  streamlines representing the gas flow
  pattern. \response{The white dashed circle indicates the
    characteristic standoff distance,
    $R_0\approx [\dot{m} v_{\rm
      src}/(4\pi\rho_0\,||\mathbf{v}_{\rm g}||^2)]^{1/2}$,
    where the total pressure of the outflow balances that of
    the incoming ambient medium
    \citep{2020MNRAS.492.2755G,2020MNRAS.494.2327L}.}}
  \label{fig:Basic-rho}     
\end{figure*}

\begin{figure} 
  \centering    
  \includegraphics[width=1.0\columnwidth]
  {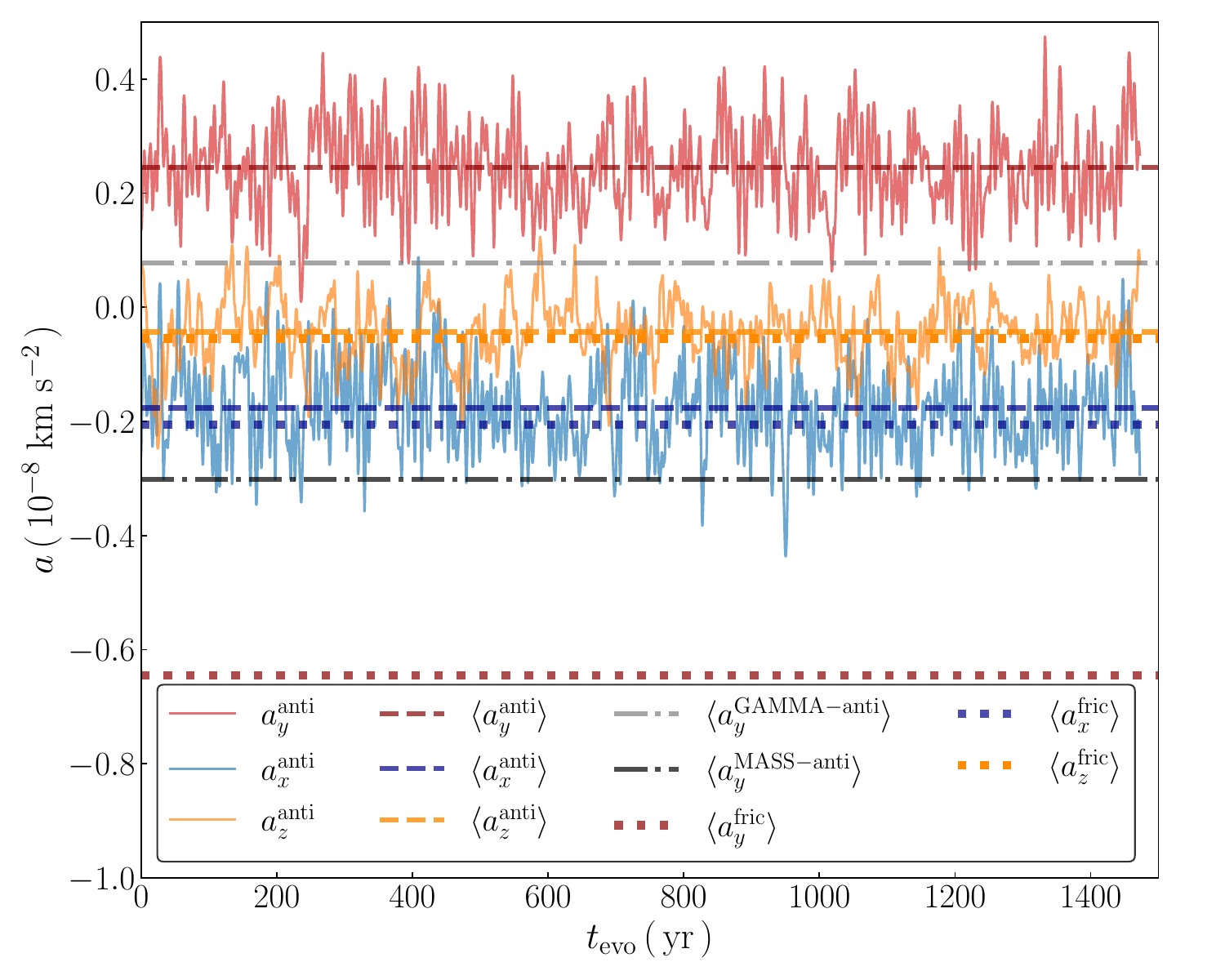}
  \caption{The acceleration experienced by the star in the
    \response{FID-anti} and FID-fric models. The red, blue,
    and orange lines represent the acceleration components
    in the $x$, $y$, and $z$ directions, respectively. Solid
    and dashed lines correspond to the \response{FID-anti}
    model, while the dotted lines represent the FID-fric
    model. \response{The grey and black dash-dot lines
      represent the GAMMA-anti and MASS-anti models. The
      solid lines depict the instantaneous acceleration,
      whereas the other lines indicate the mean values over
      15 orbital periods.}}
  \label{fig:Basic-a}     
\end{figure}

\begin{figure} 
  \centering    
  \includegraphics[width=1.0\columnwidth]
  {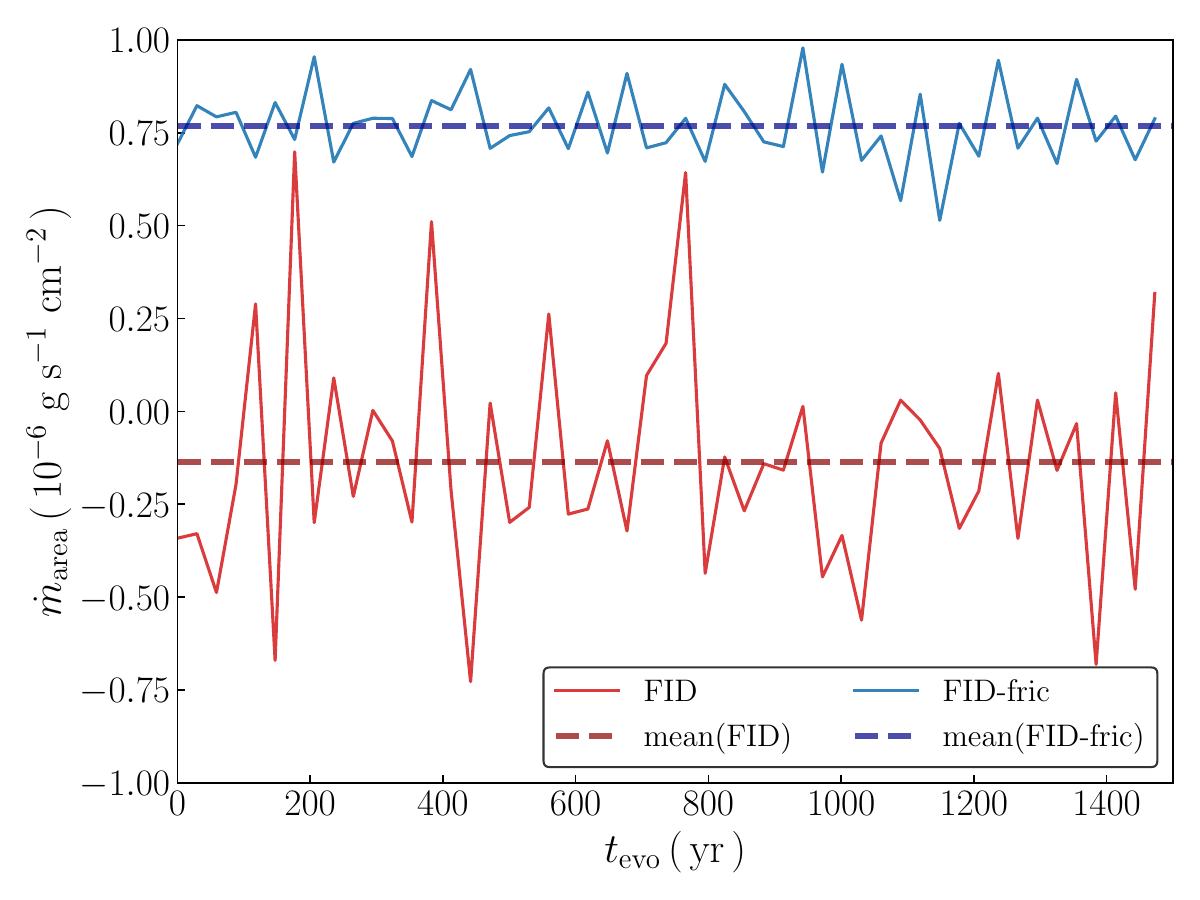}
  \caption{The mass accretion rate per area at $x =
    -L_x$. The red and blue lines correspond to the \response{FID-anti} and
    FID-fric models, respectively. The solid lines depict
    the instantaneous $\dot{m}_{\rm area}$, while the dashed
    lines represent the mean values averaged over 15 orbital
    periods.}
  \label{fig:Basic-mdot}     
\end{figure}

\begin{figure} 
  \centering    
  \includegraphics[width=1.0\columnwidth]
  {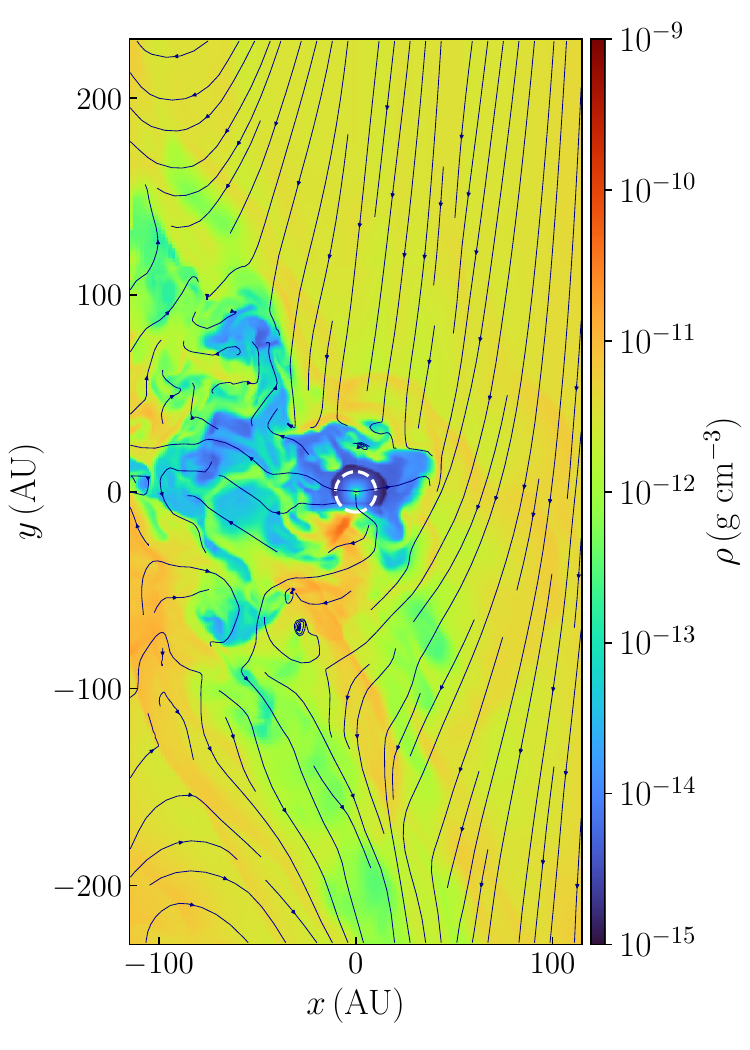}
  \caption{\response{Similar to Figure~\ref{fig:Basic-rho}, but for
    the MASS-anti model.}}
  \label{fig:Basic-massive}     
\end{figure}

\section{Methods}
\label{sec:method}

This work uses the GPU-optimized higher-order Godunov code
\texttt{Kratos} \citep{wang2025}, in simulation domains that
consider shearing flows according to Keplerian motion, to
study the dynamical evolution of a star embedded in an AGN
disk. Unless otherwise specified, all geometric-related
physical quantities are defined in the shearing box
reference frame. The coordinate system is Cartesian
($x,\,y,\,z$), with $x$ being radial, $y$ being azimuthal,
and $z$ aligned with the AGN disk's rotation axis. The star
is placed at the origin, while the SMBH is located on the negative $x$-axis at the disk
center. The masses of the star and the SMBH are $m$ and $M$,
respectively, and both lie in the plane $z=0$.

\subsection{Dynamical Equations in the Shearing
Box}\label{subsec:dynamical-eq}

As mentioned earlier, the local shearing box approximation
uses a reference frame at a radius $r_0$ (i.e., the location
of the star), co-rotating with the disk at the orbital
frequency $\Omega(r_0)=\Omega_0=\sqrt{GM/r_0^3}$
\citep{2010ApJS..189..142S}. In this frame, the hydrodynamic
(HD) equations are expressed in a Cartesian coordinate
system ($x,\,y,\,z$) with unit vectors $\hat{i}$, $\hat{j}$,
$\hat{k}$:
\begin{equation}
  \label{eq:method-HDequations}
  \begin{split}
  \partial_t \rho + \nabla \cdot (\rho \mathbf{v}) &= 0 ;
  \\
  \partial_t (\rho \mathbf{v}) + \nabla \cdot (\rho
    \mathbf{v} \mathbf{v} + p) &=  
  \rho \Omega_0^2 (2qx \hat{i} - z \hat{k})
  \\
  &- 2 \Omega_0 \hat{k} \times (\rho \mathbf{v})- \rho
    \nabla\Phi_{\rm s} - \rho f_{\rm acc}\hat{j}; 
  \\
  \dfrac{\partial E}{\partial t} + \nabla \cdot
    \left[\mathbf{v} (E + p) \right] &= 
\Omega_0^2 \rho \mathbf{v} \cdot (2qx \hat{i} - z \hat{k}) 
  \\
  &- \rho\mathbf{v} \cdot (\nabla\Phi_{\rm s}+f_{\rm
    acc}\hat{j}), 
  \end{split}
\end{equation}
where $\rho,\,\mathbf{v},\,p$ are gas density, velocity, and
pressure, respectively. The shear parameter
  $q \equiv -\d\ln\Omega/\d\ln r$ takes
value $q=3/2$ for Keplerian rotation. Total
energy density $E$ is related to the pressure $p$ and
kinetic energy density by,
\begin{equation}
  \label{eq:method-totalenergy}
   E = \dfrac{p}{\gamma - 1} + \dfrac{1}{2} \rho v^2,
\end{equation}
where $\gamma$ is the ratio of specific heats, and we take
$\gamma = 5/3$ throughout this work. The tangential
  acceleration $f_{\rm acc}$ is responsible of angular
  momentum removal that is associated with accretion
  processes, related to the radial accretion velocity
  $v_{\rm acc}$ via $f_{\rm acc} = -2\Omega_0 v_{\rm acc}$.
  Such parametrization focuses this work on the fundamental
  gas–star interactions without the complexity of detailed
  mechanims that drive accretion.  The gravitational
  potential of the star (not to be confused with the
  effective potential regarding the motion circulating the
  central blackhole) is $\Phi_{\rm s}$.  Such a central star
  is treated as a source particle (see e.g.,
  \citealt{2017MNRAS.465.1316M}) that launches outflows.  In
  order to consistently resolve the flow outside the source
  region ($r_{\rm src} < r \leq r_{\rm soft}$) without
  excessively reducing the time step, the gravitational
  potential of the perturber is softened
  \citep{2011ApJ...741...56D}. Within this range, a
  fourth-order gravitational potential is applied,
\begin{equation}
  \label{eq:method-gravity4th}
   \Phi_{\rm s}^{(4)} = -G M_{\rm s} \dfrac{r^2 + 1.5r_{\rm
       soft}^2}{\left( r^2 + r_{\rm soft}^2 \right)^{3/2}}. 
\end{equation}
This work adopts $r_{\rm soft} / r_{\rm src} \approx 3$
throughout all simulations. Following the approach in
\citet{2022ApJ...932..108W}, a spherical source region with
a radius $r_{\rm src}$ is established around the star. When
considering stellar outflows, an initial radial velocity
$v_{\rm src}$ is set within the source region as a model
parameter, and the density is adjusted to produce the
desired mass-loss rate
$\dot{m} = 4\pi r^2 \rho v_{\rm src}$, where
$r^2=x^2+y^2+z^2$. The gas pressure is determined based on a
typical temperature $T_{\rm outflow}$. During the dynamical
evolution, the physical quantities are fixed within the
source region, ensuring constant outflows.

\subsection{Basic Disk Model and Boundary Conditions}
\label{subsec:disk-bc}

The 3D equilibrium AGN disk model is considered
\citep{2013MNRAS.435.2610N}.  Since the validity of the
shearing box approximation requires that the domain size be
small relative to $r_0$ \citep{1995ApJ...440..742H}, the
initial gas density $\rho$ and pressure gradient
$\partial p/ \partial x$ are assumed to be uniform in the
$x$-direction, with a constant temperature $T$ throughout
the shearing box. The pressure gradient in the $y$-direction
is assumed to be zero. The self-gravity of the gas is also
neglected in this model, which is safe for the study
  focusing on the vincinity of stars embedded in AGN disks
  \citep[see e.g.][]{2022MNRAS.517.1602L}.

Along the $z$-direction, the pressure gradient
  balances the gravitational force from the central SMBH,
\begin{equation}
  \label{eq:method-balance-zeq}
  \begin{split}
    - \dfrac{\partial p}{\partial z} - \rho \Omega_0^2 z =
    0\ ,
    p = \dfrac{k_{\rm B}T}{\mu m_{\rm p}} \rho\ ,
  \end{split}
\end{equation}
where $m_{\rm p}$ is the proton mass, $\mu=2.35$ is
  the mean molecular weight measured in $m_{\rm p}$, and
  $k_{\rm B}$ is the Boltzmann constant. Solution to the
vertical force balance yields,
\begin{equation}
  \label{eq:method-rho-dis}
    \rho = \rho_0\,\exp\left(-\dfrac{z^2}{2H^2_0}\right),
\end{equation}
where $\rho_0$ is the gas density at the $z=0$ plane,
$H_0 = c_{\rm iso}/\Omega_0$ is the disk scale height, and
$c_{\rm iso}^2=p/\rho$ is the isothermal sound speed. The
disk-only steady state of eq.\eqref{eq:method-HDequations}
reads, in absence of the star,
\begin{equation}
  \label{eq:method-veldisk}
  \mathbf{v}_{\rm g} = -q\Omega_0 (x+x_{\rm p})\hat{j}\ ,\ 
  x_{\rm p} = -\dfrac{\partial_x p}{2q\Omega_0^2 \rho}\ ,
\end{equation}
which describes the uniform orbital motion of gas in the
shearing box. We further introduce the parameter $x_{\rm p}$
for the offset between the orbiting body and the disk gas
that moves with the same orbital velocity, caused by the
presence of a radial pressure gradient in the disk
\citep{2017MNRAS.472.4204M}. For typical pressure profiles
that decrease with radius, $x_{\rm p} > 0$ indicates that
the tangential velocity of the object is greater than the
ambient gas, resulting in a headwind.

\response{More consistent thermodynamics is also a direction
  of potential improvements.  This work assumes that the
  heating introduced by stellar outflows is negligible
  compared to the radiative cooling of the AGN disk. Taking
  the fiducial model as an example, the heating rate ratio
  $\eta$ is estimated as
\begin{equation}
  \label{eq:heating-rate}
    \begin{split}
    Q^{+}_{\rm kin} &\approx \dfrac{\dot{m}v^2_{\rm src}}{2}\times\dfrac{1}{4\pi R^2_0}=\dfrac{\rho_0||\mathbf{v}_{\rm g}||^2}{2}v_{\rm src},
    \\
    Q^{-}_{\rm rad} &\equiv 2\sigma T^4_{\rm e}=2\sigma T^4\dfrac{\tau}{1+\tau^2},
    \\
    \eta &= \dfrac{Q^{+}_{\rm kin}}{Q^{-}_{\rm rad}}\approx10\%\times\left( \dfrac{\rho_0}{3.5\times10^{-11}~{\rm g~cm^{-3}}}\right)
    \\
    &\times \left( \dfrac{||\mathbf{v}_{\rm g}||}{4.1\times10^{6}~{\rm cm~s^{-1}}}\right)^{2}\times \left( \dfrac{v_{\rm src}}{8\times10^{7}~{\rm cm~s^{-1}}}\right)
    \\
    &\times \left( \dfrac{T}{2.8\times10^{4}~{\rm K}}\right)^{-4},
    \end{split}
\end{equation}
where
$R_0\approx [\dot{m} v_{\rm
  src}/(4\pi\rho_0\,||\mathbf{v}_{\rm g}||^2)]^{1/2}$,
$Q^{-}_{\rm rad}$ is the cooling rate due to radiative
diffusion in the direction normal to the disk and
$\tau \approx H_0 \kappa_{\rm es} \rho_0$ is the Rosseland
mean opacity \citep{1990ApJ...351..632H,2003ApJ...597..131J,
  2011ApJ...730...45J,2021ApJ...910...94C}. The above
calculations show that the kinetic heating from stellar
outflow is insufficient to offset radiative cooling and
maintain thermal equilibrium, even though turbulent heating
alone has been found inadequate in the outer regions of AGN
disks \citep{2003MNRAS.341..501S,2005ApJ...630..167T}.}

The size of the simulation box in this study is given by
$(-L_x, L_x) \times (-L_y, L_y) \times (-L_z, L_z)$, where
$L_x,L_y,L_z$ represent the lengths in the
$x,y,z$-directions, respectively. The BCs for the shearing
box are set similarly to those in
\citet{2011ApJ...741...56D}. At the $x$-boundaries, the
physical variables in the ghost zones are fixed at their
unperturbed Keplerian values at $x=\pm L_x$ (i.e.,
maintained in their initial state).
Both sides of the $x$-boundaries allow waves to exit upon
reaching the edges. For the $y$-boundaries, an
inflow/outflow BC is applied. Variables in the ghost zones
are fixed at their initial values at physical "inflow"
boundaries ($y<0,\,x+x_{\rm p}<0$ and
$y>0,\,x+x_{\rm p}>0$), and are copied from the last
actively updated row of cells at physical "outflow"
boundaries ($y<0,\,x+x_{\rm p}>0$ and
$y>0,\,x+x_{\rm p}<0$). For the $z$-boundaries, a
conventional outflow BC is used at $z=\pm L_z$.

\begin{figure*} 
  \centering    
  \includegraphics[width=0.95\columnwidth]
  {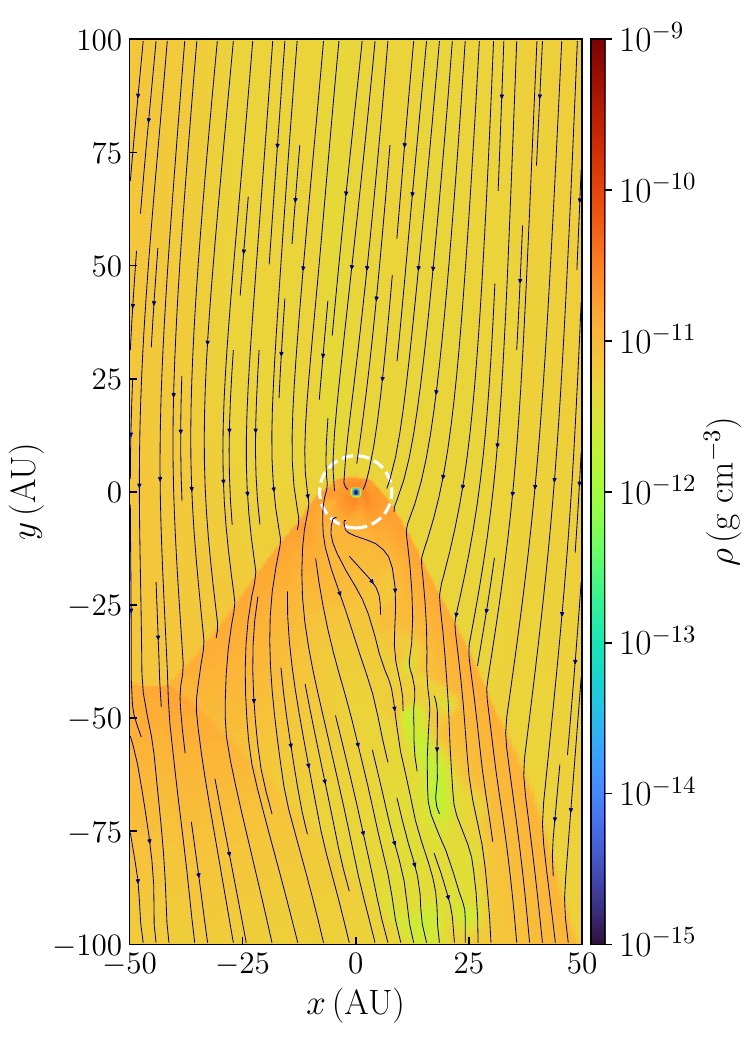}
  \includegraphics[width=0.95\columnwidth]
  {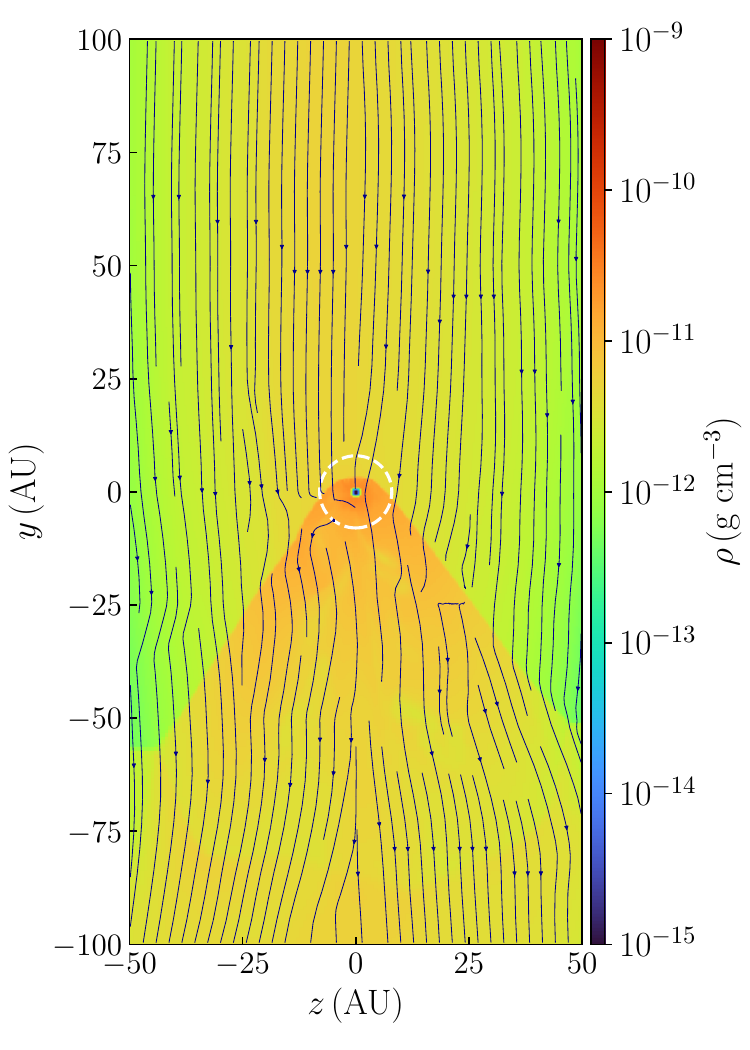}
  \caption{Similar to Figure~\ref{fig:Basic-rho}, but for
    the star without outflow in the fiducial
    model. \response{The white dashed circle indicates the
      Bondi radius,
      $R_{\rm B} = 2Gm/(c_{\rm iso}^2 + ||\mathbf{v}_{\rm
        g}||^2)$.}}
  \label{fig:Basic-f-rho}     
\end{figure*}

\subsection{Setup of Fiducial and Other Models}
\label{subsec:setup}

The fiducial model contains an $8~M_\odot$ star embedded in
the outer regions of an AGN disk, orbiting a central SMBH
with a mass of $10^8~M_\odot$. The star generates isotropic
outflows that interact with the ambient gas. For comparison,
the star without outflows is also considered. The initial
disk parameters and the definition of the outer region are
adopted according to \citep{2021ApJ...910...94C}, with the
outer region defined as the area where
$ Q \equiv M/\big(\sqrt{8}\pi\rho r^3\big) \lesssim 1$. In
the fiducial model, the star is placed at a typical distance
of $r_0 = 5000~R_{\rm sch}\approx0.05~{\rm pc}$, where
$R_{\rm sch}\equiv2GM/c^2$ is the Schwarzschild radius and
$c$ is the speed of light. \response{It is noted that such
  radius approximately corresponds to the critical boundary
  separating the regions 
  dominated and radiation-pressure-dominated
  regions \citep{2021ApJ...910...94C}. The star is assumed
  to reside in a gas-pressure-dominated region throughout
  this work.}  Other specific parameter settings are
provided in Table~\ref{tab:fiducial-property}.

Stars in AGN disks could be very massive with intense
  nuclear burning, many of which could reach the Eddington
  luminosity $L_{\rm Edd}$ \citep{2021ApJ...910...94C}. When
  the stellar luminosity exceeds the Eddington limit,
  intensive mass-loss is expected
\citep[e.g.,][]{2012ASSL..384..275O,
  2014ARA&A..52..487S}. Following
\citet{2011ApJS..192....3P}, this work assumes a
super-Eddington outflow at the escape velocity
$v_{\rm esc}$,
\begin{equation}
  \label{eq:method-masslossrate}
    \begin{split}
    \dot{m}&=\dfrac{L_{\rm Edd}}{v_{\rm esc}^2},
    \\
    L_{\rm Edd} &\equiv -\dfrac{4\pi Gmc}{\kappa},z
    \\
    &\approx2.56\times10^5\left(\dfrac{\kappa}{\kappa_{\rm
      es}}\right)^{-1}\left(\dfrac{m}{8~M_\odot}\right)L_\odot, 
    \\
    v_{\rm esc}&=\left(\dfrac{2Gm}{r_{\rm s}}\right)^{1/2},
    \end{split}
\end{equation}
where $\kappa$ is the opacity, with the electron-scattering
$\kappa_{\rm es}$ value adopted. $r_{\rm s}$ is the stellar
radius, which follows the approximate scaling relation
$r_{\rm s} \propto m^{0.8}$ \citep{1991Ap&SS.181..313D}. The
temperature of the stellar outflow is estimated based on the
star's surface temperature \citep{2021ApJ...910...94C}. The
outflow temperature, $T_{\rm outflow}$, is expressed
as: \begin{equation}
  \label{eq:method-t-outflow}
    T_{\rm outflow} \equiv \left( T^4 + \dfrac{L_{\rm
          Edd}}{4\pi r_{\rm s}^2 \sigma} \right)^{1/4}, 
\end{equation}
where $\sigma$ is the Stefan-Boltzmann constant.

In addition to the fiducial model, cases with different
pressure gradients are explored in
\S\ref{sec:pressure-gradient-effect}, anisotropic
outflows (``jets'') are studied in \S\ref{sec:jet-effect},
and \S\ref{sec:accretion-disk} and \S\ref{sec:black-hole}
study the specific cases where stars and sBHs are tested in
accreting AGN disks. All models are summarized in Table~\ref{tab:model-summary},
characterizing each model with the prescribed dynamical
evolution time $t_{\rm evo}$, the time-averaged acceleration
along the $y$-axis $\overline{a_y}$ under the effect of
anti-friction, and the corresponding mean rate of change in
orbital radius $\dot{r}_{\rm cir}$. Following the approach
of \citet{2011ApJ...741...56D}, all the simulations run for
several extra periods $P$ before measuring the dynamical
properties, excluding the influence of time-dependent
structures on the star.

\begin{figure*} 
  \centering    
  \includegraphics[width=0.95\columnwidth]
{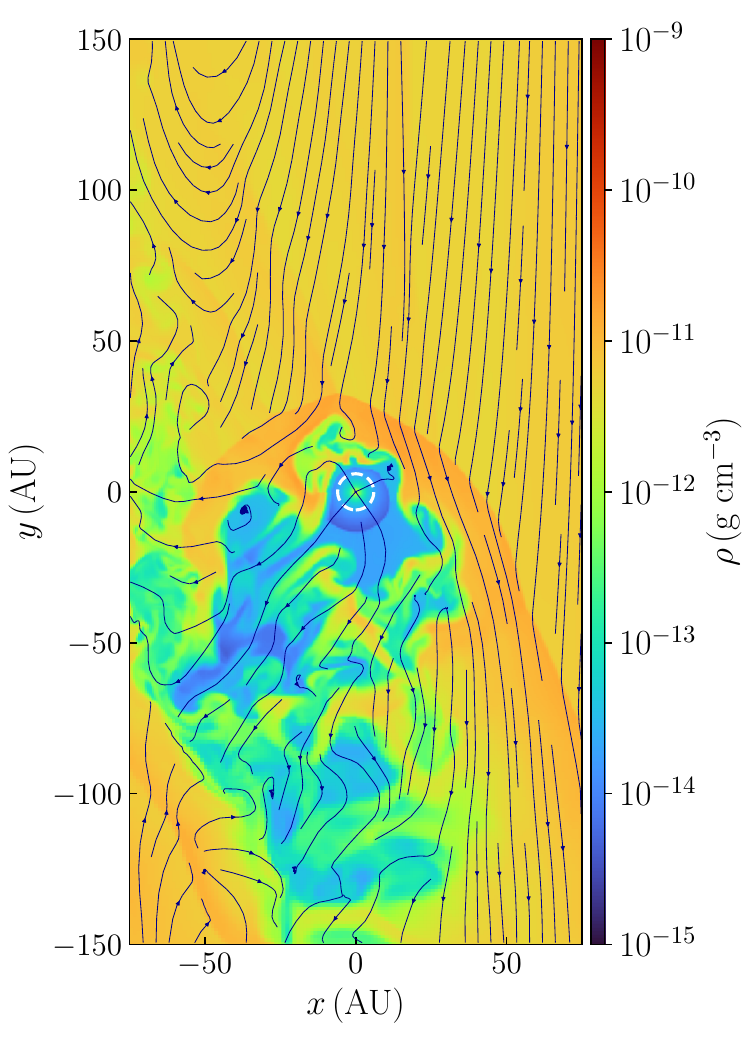}
  \includegraphics[width=0.95\columnwidth]
{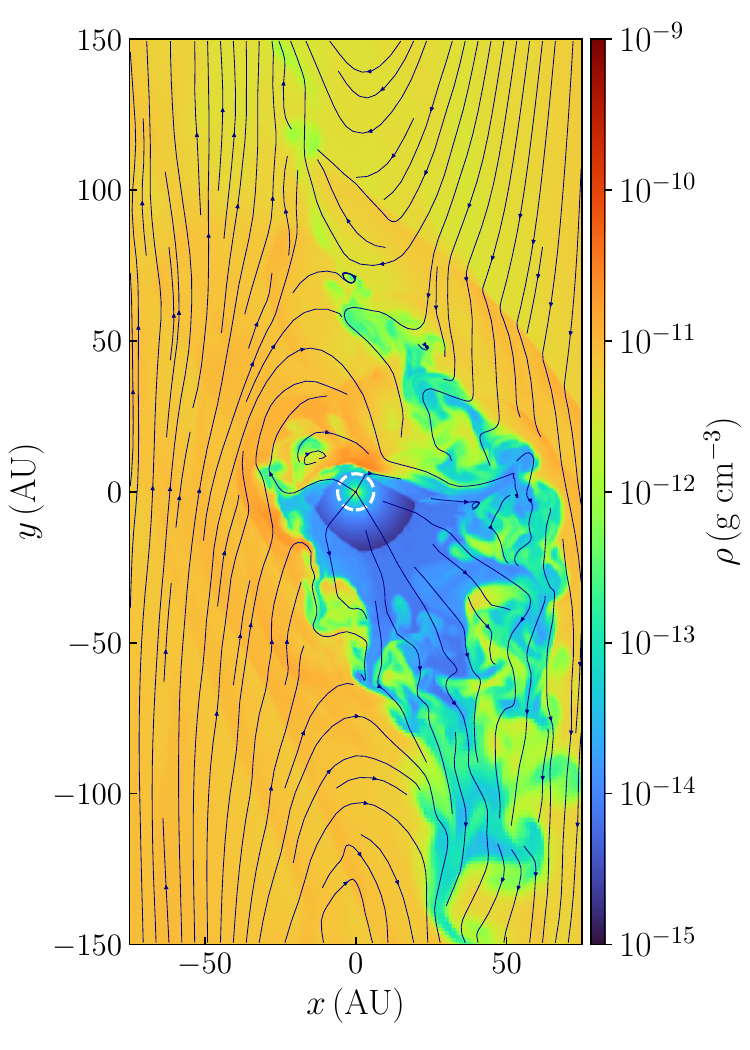}
  \caption{Gas density $\rho$ in the $z=0$ plane (i.e., the
    orbital plane) at $t_{\rm evo} = 15~P$ for the PG-45
    (left panel) and PG-0 (right panel) models. \response{The white dashed circle indicates the characteristic standoff distance $R_0$.}}
  \label{fig:PG-45-0-rho}     
\end{figure*}

\section{Fiducial Model Studies}
\label{sec:result-fiducial}

\subsection{Gas-Star Interaction and Angular Momentum
  Transfer} 
\label{sec:gas-star-interaction}

  Figure~\ref{fig:Basic-rho} illustrates the
  interactions between the star and disk gas for the 
  fiducial model \response{FID-anti}, showing the density profile on the
  $z=0$ and $x=0$ planes at $t_{\rm evo}=15~P$. The pressure
  gradient leads to sub-Keplerian motion in the
  $x+x_{\rm p}>0$ region, resulting in a headwind acting on
  the star. Combined with the star's outflow, this
  interaction gives rise to a bow shock structure on the
  leading side of the orbital motion, whose overdense in
  front of the moving star could lead to the anti-friction
  effect. 


  Figure~\ref{fig:Basic-a} shows the time dependence
  of all three components of the star's acceleration, with
  dashed lines representing the mean values. The time
  average of the vertical component $a_z$ is very close to
  zero, which is expected due to the reflection symmetry of
  the system.  The $a_y$ component, which dominates the
  orbital evolution due to angular momentum transfer, is
  positive because of the anti-friction. In the long run,
  the star is accelerated tangentially, moving outwards due
  to higher specific angular momentum caused by
  anti-friction. The asymmetry bow shock over the
  $x$-direction also induces a gravitational pull toward the
  central SMBH, yet the long-term effect of radial
  acceleration is negligible compared to tangential
  motion. Based on the Gauss planetary equations and the
  shearing box approximation, the evolution of the orbital
  radius $r_{\rm cir}$ reads,
\begin{equation}
  \label{eq:r_time}
  \dot{r}_{\rm cir} \approx \dfrac{2}{\Omega_0} a_y\ .
\end{equation}
  Quantitatively,
  $\langle\dot{r}_{\rm cir}\rangle \approx 2.4~{\rm km}~{\rm s^{-1}} > 0$.
  Under the influence of anti-friction, the angular
  acceleration causes $1\%~v_0$ (where $v_0 = \Omega_0 r_0$)
  acceleration within approximately $400~{\rm yr}$. 
  Such acceleration is ultimately caused by the
  extraction of angular momentum from the disk gas, which
  should result in the enhancement of disk gas accretion.
  Figure~\ref{fig:Basic-mdot}
  presents the mass accretion rate per unit area
  $\dot{m}_{\rm area}$ at $x=-L_x$, with the 15-period
  average being approximately
  $-10^{-7}~{\rm g}~{\rm s}^{-1}~{\rm cm}^{-2}$ (where the
  negative value indicates accretion).

The case without stellar outflows is studied as the
controlled group, keeping all other parameters unchanged,
and is denoted as the FID-fric model. To reduce
computational cost, the simulation box is reduced to
$(L_x,L_y,L_z) =
(50,\,100,\,150)~\rm{AU}$. Figure~\ref{fig:Basic-f-rho}
presents the density profile at $t_{\rm evo}=15~P$ for the
FID-fric. Similar to the \response{FID-anti} model, the density wave remains
symmetric about $z=0$ but asymmetric about $x=0$. However,
in this case, gas accumulates behind the moving star,
forming an overdense wake that exerts a drag force, leading
to dynamical friction. The dotted line for $a_y$ in
Figure~\ref{fig:Basic-a} presents the acceleration of the
star under dynamical friction, taking the opposite sign
compared to the anti-friction case. Such comparison confirms
that the anti-friction effect can flip the direction of
orbit migration of stars embedded in AGN disk scenarios
resulting in inward migration toward the disk center
at a rate
$\langle\dot{r}_{\rm cir}\rangle \approx -6.3~{\rm km}~{\rm s^{-1}}$.
In the meantime, the ambient gas gains angular momentum and
the accretion is inhibited and even reverted as a decretion
flow, as indicated by Figure~\ref{fig:Basic-mdot}. 

\response{Given that massive stars tend to yield higher
  mass-loss rates, an additional run simulates an
  $80\,M_\odot$ star, referred to as the MASS model.  This
  case sets the outflow velocity to
  $v_{\rm src} = 2\times10^8~{\rm cm~s^{-1}}$, while other
  parameters remain identical to the fiducial model. As
  shown in Figures~\ref{fig:Basic-a} through
  \ref{fig:Basic-massive}, the massive star experiences
  deceleration in the $y$-direction, in contrast to the
  acceleration observed in the FID-anti model. This
  difference arises mainly from the stronger gravitational
  concentration in the wake of the massive star on the
  ambient gas and the outflows outside the terminal shock,
  which leads to a more chaotic interaction between the
  outflow and the ambient gas that is dominated by the
  over-dense region on the trailing side.
}


\begin{figure} 
  \centering    
  \includegraphics[width=1.0\columnwidth]
{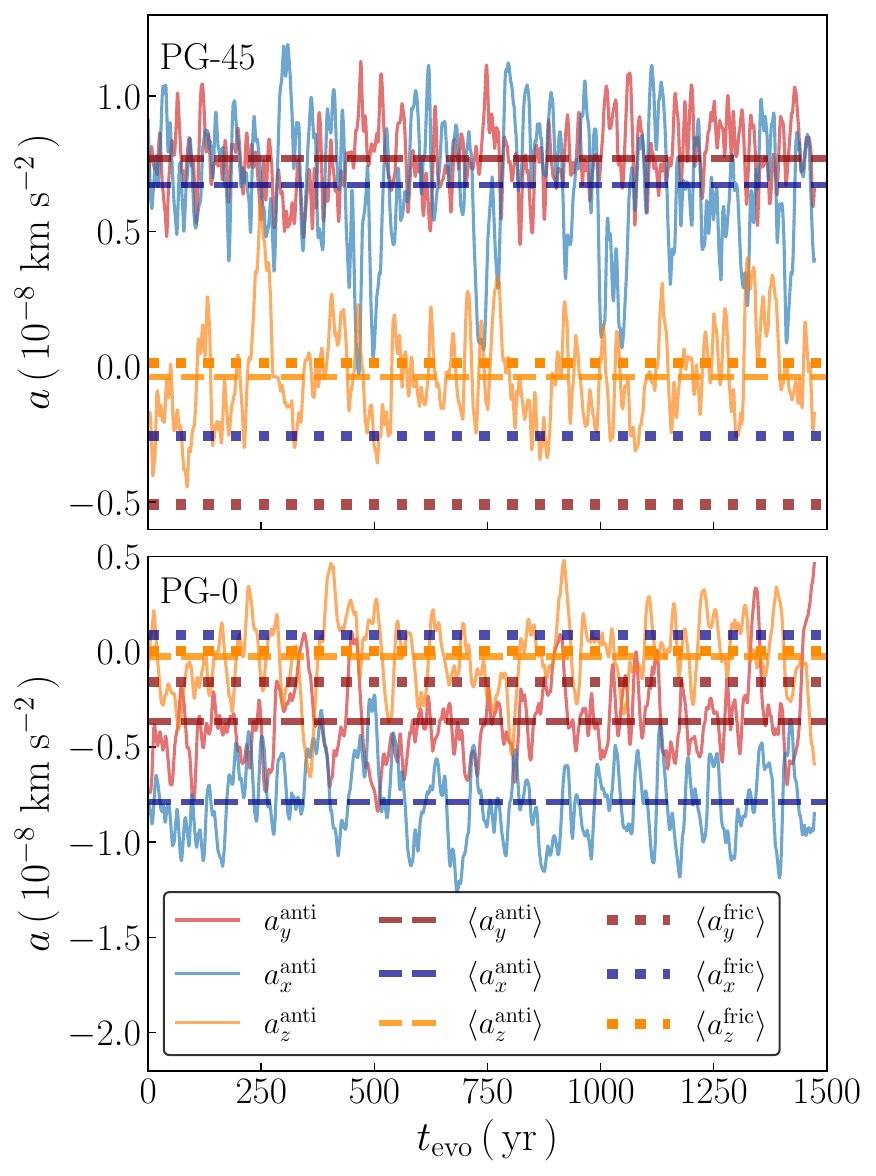}
\caption{The acceleration experienced by the star in the
  PG-45 (upper panel) and PG-0 (lower panel) models. Both
  models include scenarios with and without stellar
  outflows.}
  \label{fig:PG45-0-a}     
\end{figure}

\begin{figure} 
  \centering    
  \includegraphics[width=1.0\columnwidth]
{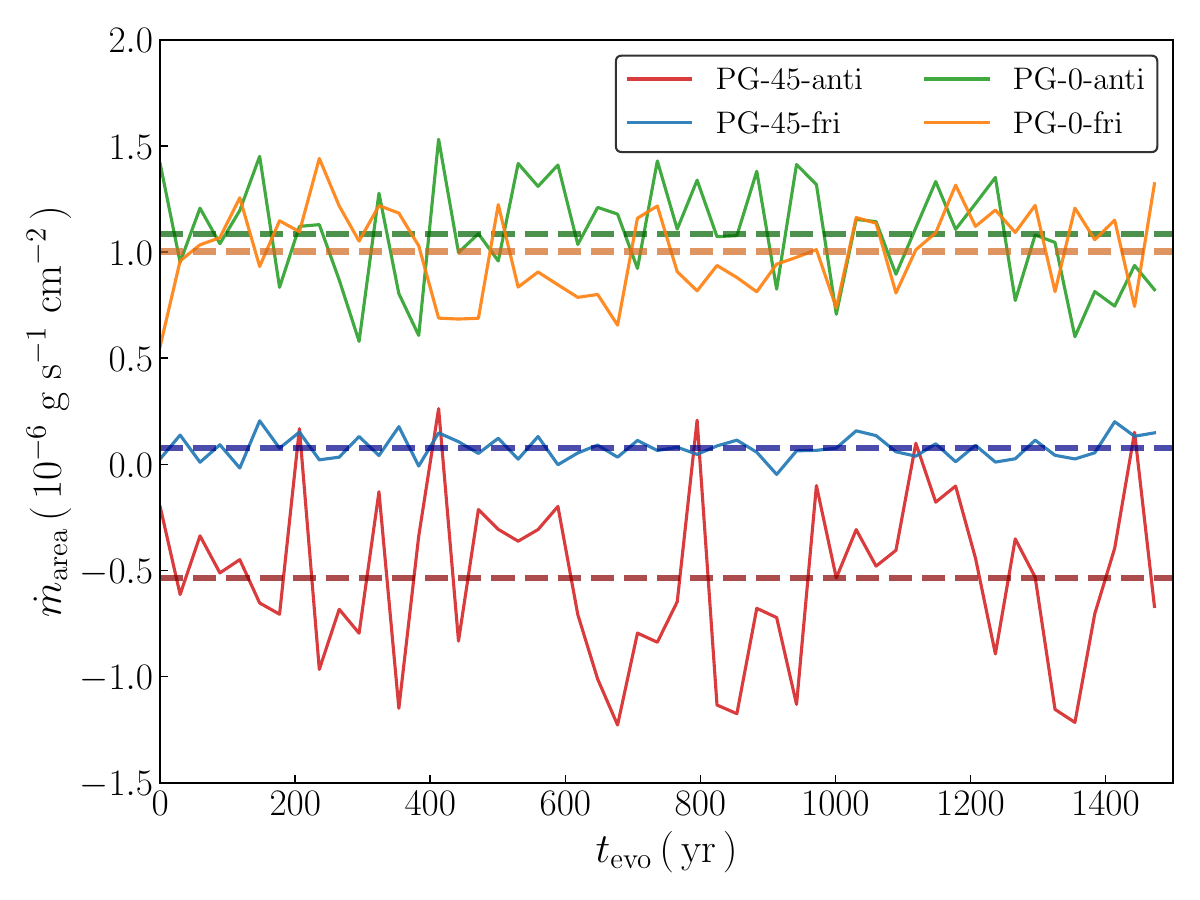}
\caption{Similar to Figure~\ref{fig:Basic-mdot}, but for the
  PG-45 and PG-0 models.}
  \label{fig:PG45-0-mdot}     
\end{figure}

\begin{figure} 
  \centering    
  \includegraphics[width=1.0\columnwidth]
  {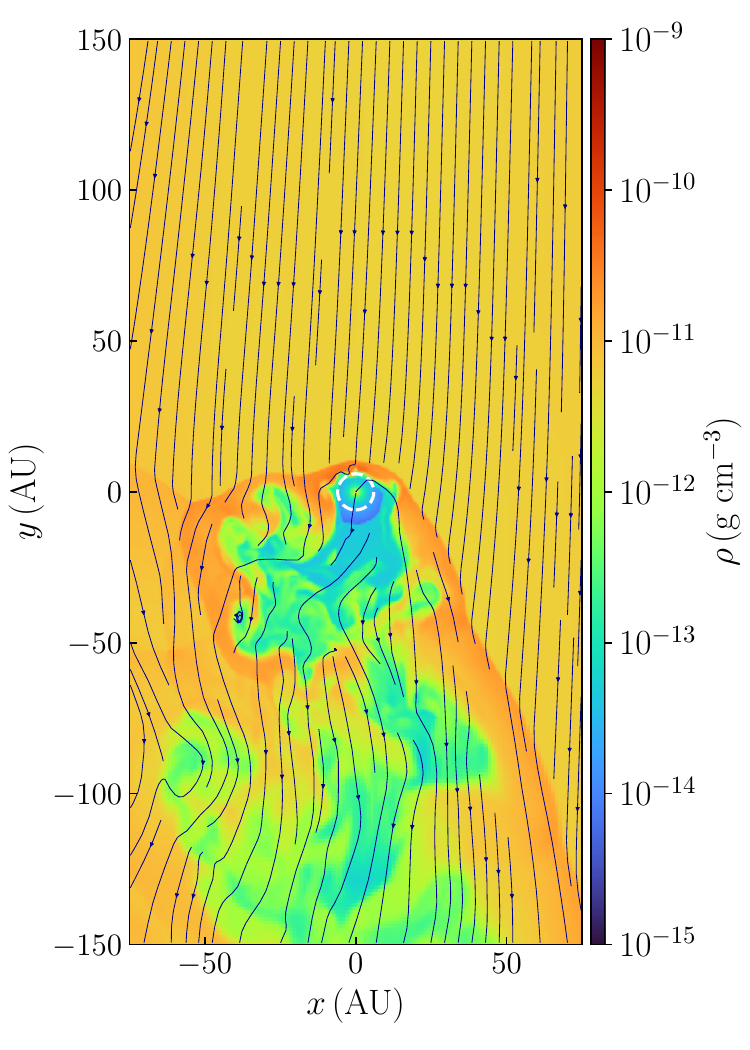}
  \caption{\response{Similar to Figure~\ref{fig:Basic-rho}, but for
    the GAMMA-anti model.}}
  \label{fig:Basic-gamma}     
\end{figure}

\subsection{Pressure Gradient}
\label{sec:pressure-gradient-effect}

  Both the \response{FID-anti} and FID-fric models identify that the
  radial distribution of tangential ($y$-component) velocity
  is key to the angular momentum acquisition.  Because of
  the radial balance of force in the co-rotating frame
  (eq.~\ref{eq:method-veldisk}), radial distribution of the
  relative gas velocity is primarily determined by the
  pressure gradient. Two additional models are hence
  introduced to explore the influence of radial pressure
  gradient, keeping all other parameters of the \response{FID-anti} model
  intact. The parameter $x_{\rm p}$ is set to $45~\rm AU$
  and $0~\rm AU$, corresponding to models PG-45 and PG-0,
  respectively, where PG-0 represents a scenario where the
  radial pressure gradient vanishes, and greater $x_{\rm p}$
  values indicates a steeper pressure gradient.

  Figure~\ref{fig:PG-45-0-rho} presents the density
  profiles in the orbital plane, showing the gas-star
  interactions of the PG-45 and PG-0 models. Compared to the
  \response{FID-anti} model, the gas density distribution in PG-45 and PG-0
  exhibits increased irregularity due to stronger velocity
  shears. The reduced pressure gradient shifts the location
  where the gas velocity matches the orbital velocity,
  weakening the head-wind structure. The star accelerations
  in the PG-45 and PG-0 models are presented in
  Figure~\ref{fig:PG45-0-a}, 
  showing distinct behaviors especially about the tangential
  component $a_y$. In the $y$-direction, the acceleration in
  PG-45 is higher than in the \response{FID-anti} model, because the
  magnitude of anti-friction is roughly inversely
  proportional to the relative velocity between the gas and
  the star \citep{2020MNRAS.494.2327L}. In contrast, PG-0
  exhibits significantly weaker anti-friction effect, which
  eventually yields a negative average $\langle a_y\rangle$
  even with outflows, due to the disrupted bow-shock
  structures.

  Figure~\ref{fig:PG45-0-mdot} shows that in PG-0,
  regardless of friction or anti-friction, the decretion
  onto the SMBH remains significant as the star loses
  angular momentum to the gas. PG-45, in contrast, exhibits
  consistently positive mean values for both $a_x$ and $a_y$
  due to anti-friction, indicating the opposite direction of
  angular momentum transfer.

\begin{figure*} 
  \centering    
  \includegraphics[width=2.1\columnwidth]
{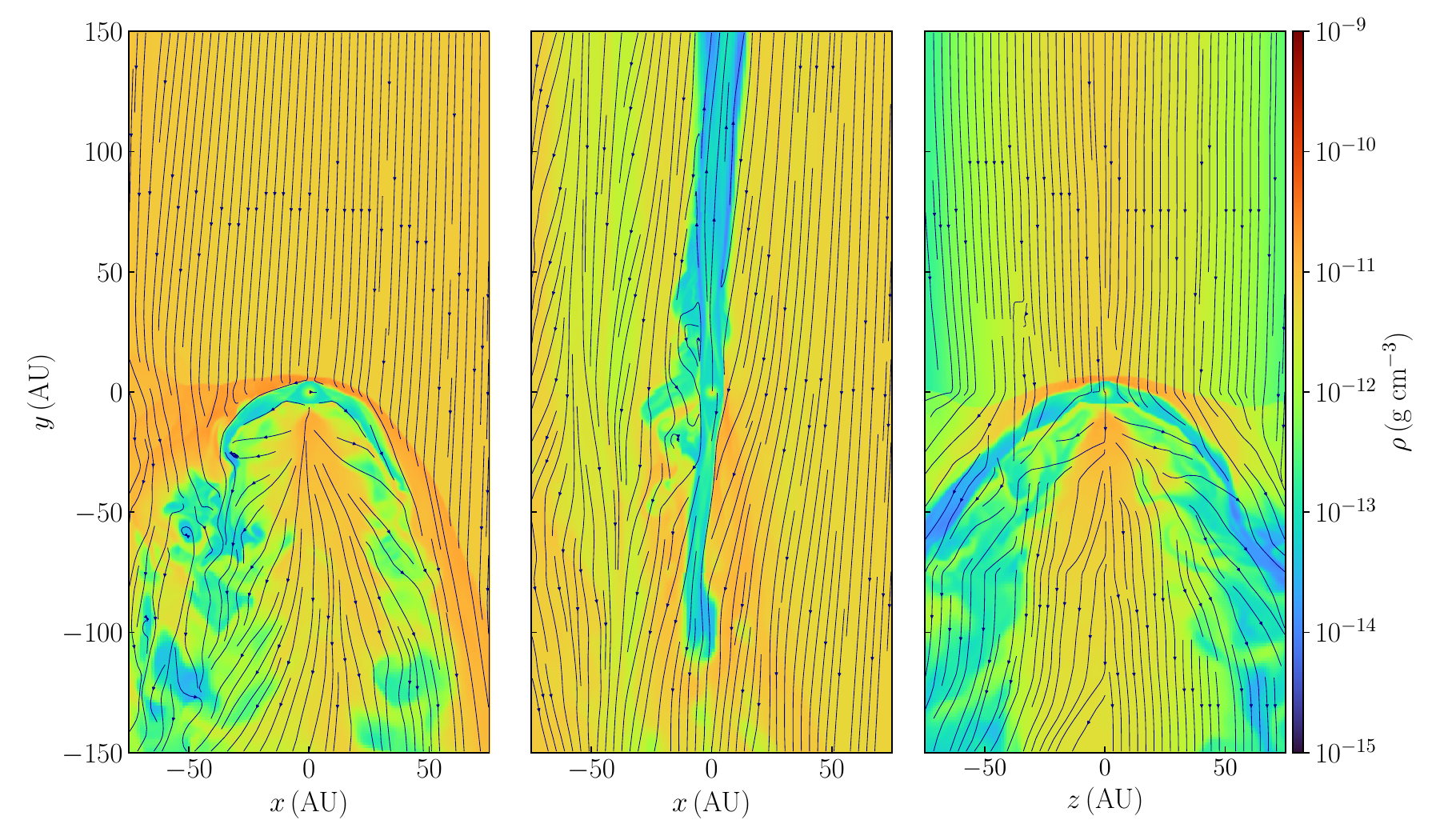}
\caption{Similar to Figure~\ref{fig:PG-45-0-rho}, but for
  the JET-x (left panel), JET-y (middle panel), and JET-z
  (right panel) models.}
  \label{fig:JET-x-y-rho}     
\end{figure*}

\begin{figure} 
  \centering    
  \includegraphics[width=1.0\columnwidth]
{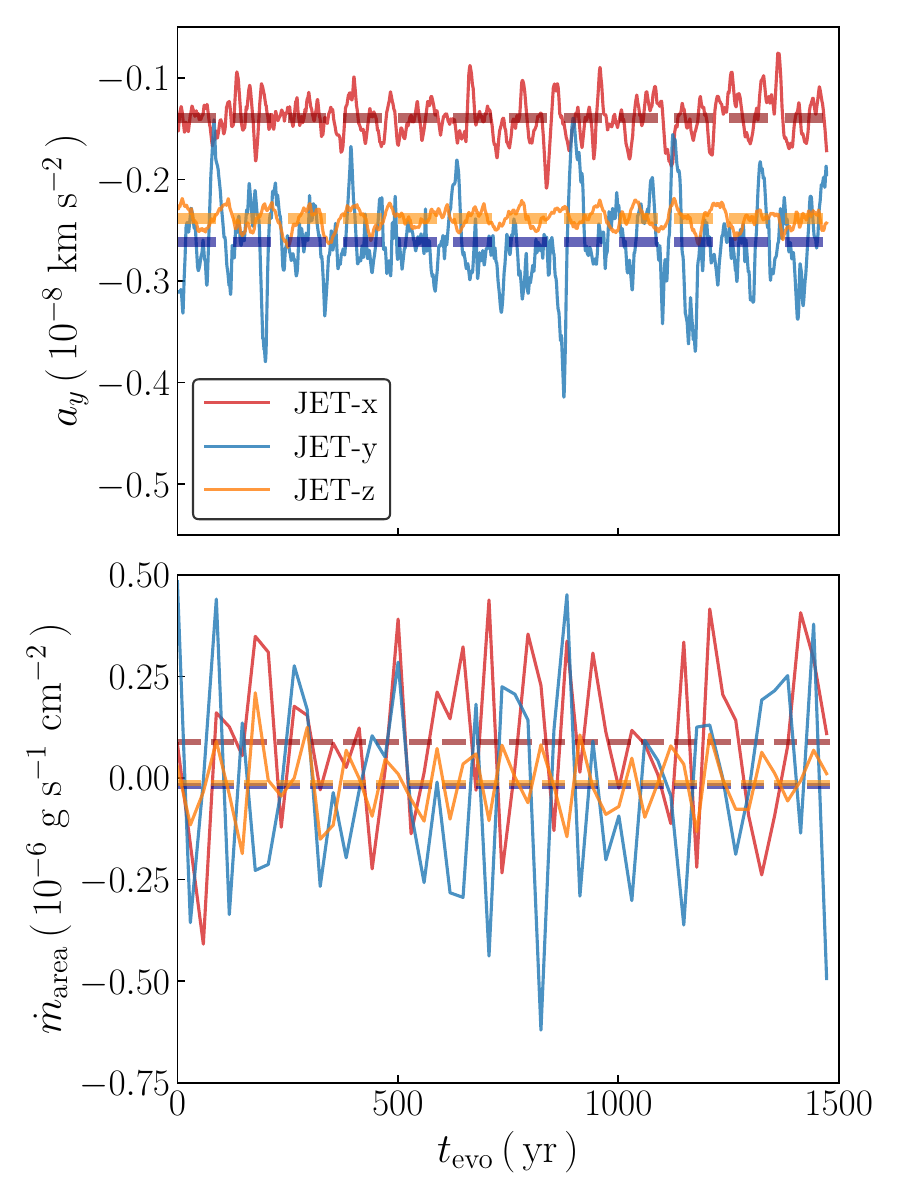}
\caption{The star's y-direction acceleration (upper panel)
  and the mass accretion rate per area at $x = -L_x$ (lower
  panel) of JET-x, JET-y, and JET-z models. Both models
  consider the scenario with stellar outflow.}
  \label{fig:JET-x-y-a-mdot}     
\end{figure}

\subsection{Anisotropic Outflow}
\label{sec:jet-effect}

  If anisotropic outflows are launched from stars, the
  pattern of gas-star interactions might be significantly
  modified. Extra models are explored assuming jet-shaped
  stellar outflows, keeping all other parameters of the
  \response{FID-anti} model unchanged. The jets are launched along $\pm x$,
  $\pm y$, and $\pm z$ directions, corresponding to models
  marked as JET-x, JET-y, and JET-z, respectively. 
  Jets are launched in collimation from both poles, whose
  mass flux are parameterized by
  $\dot{m} = 2\pi r^2 \rho v_{\rm src}$.

Figures~\ref{fig:JET-x-y-rho}-\ref{fig:JET-x-y-a-mdot} illustrate the density
distributions for gas-star interactions in all JET
models. The JET-x and JET-z simulations behave similarly, in
which jet outflows along the $x$ and $z$ directions create
low-density regions on both sides of the star (with respect
to $x=0$ or $z=0$), while the head-wind structure in the
$y$-direction remains comparable to that in the \response{FID-anti}
model. In both cases, the star experiences dynamical
friction in the $y$-direction, indicating inward migration
toward the disk center. In Model JET-x, the radial pressure
gradient affects the symmetry of the interaction between the
outflow and the ambient gas near $x=0$. Such gradient
suppresses the formation of the dense wake trailing the
star, resulting in a weaker decelerating effect in the
$y$-direction compared to the JET-z case. Due to the jet
launched along the $-x$ direction, the specific mass
accretion rate $\dot{m}_{\rm area}$ at $x = -L_x$ is
negative (decretion). The time-averaged $\dot{m}_{\rm area}$
in the JET-z model is approximately zero.  The JET-y model
with strong $y$-outflows prevents the formation of an
overdense region ahead of the star, effectively leading to a
decelerating force in the $y$-direction. Since gas-star
interaction in the JET-y model is primarily concentrated in
the $y$-direction, the inflow effect on the gas along the
$x$-direction is negligible, leading to an average
$\dot{m}_{\rm area}$ at $x = -L_x$ close to zero.

\begin{figure*} 
  \centering    
  \includegraphics[width=0.95\columnwidth]
  {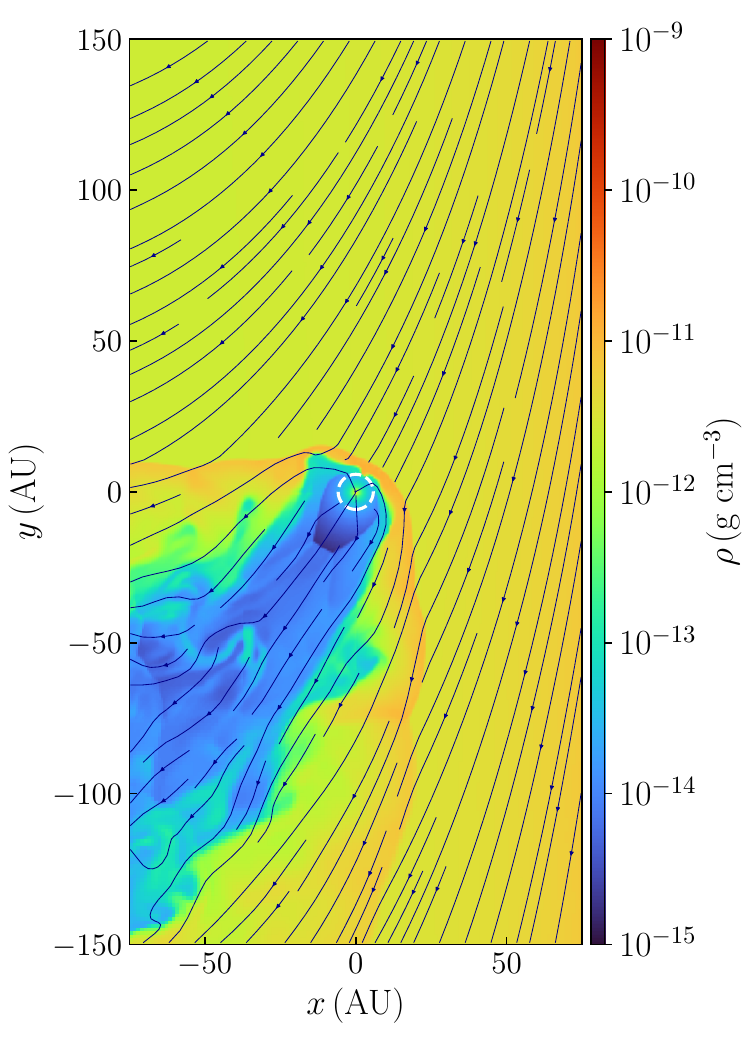}
  \includegraphics[width=0.95\columnwidth]
  {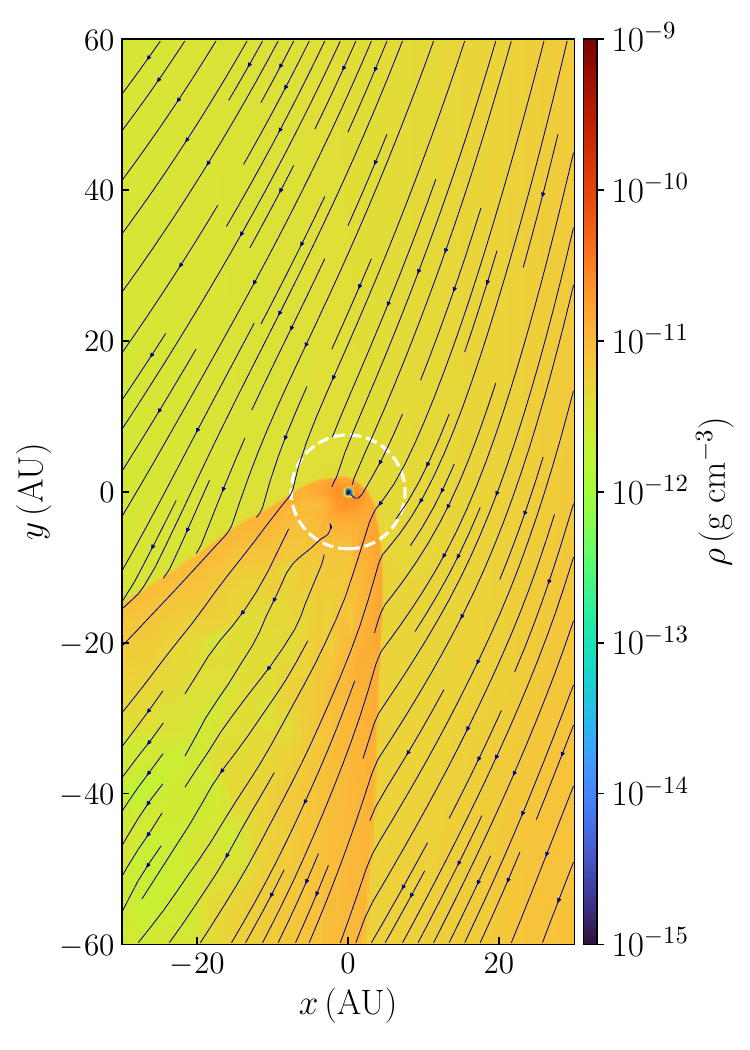}
  \caption{Similar to Figure~\ref{fig:PG-45-0-rho}, but for
    the ACCD-anti (left panel) and ACCD-fric (right panel)
    models. \response{The white dashed circles indicate the
      characteristic standoff distance $R_0$ in the left
      panel and the Bondi radius $R_{\rm B}$ on the right.}}
  \label{fig:ACCD-rho}     
\end{figure*}

\begin{figure} 
  \centering    
  \includegraphics[width=1.0\columnwidth]
  {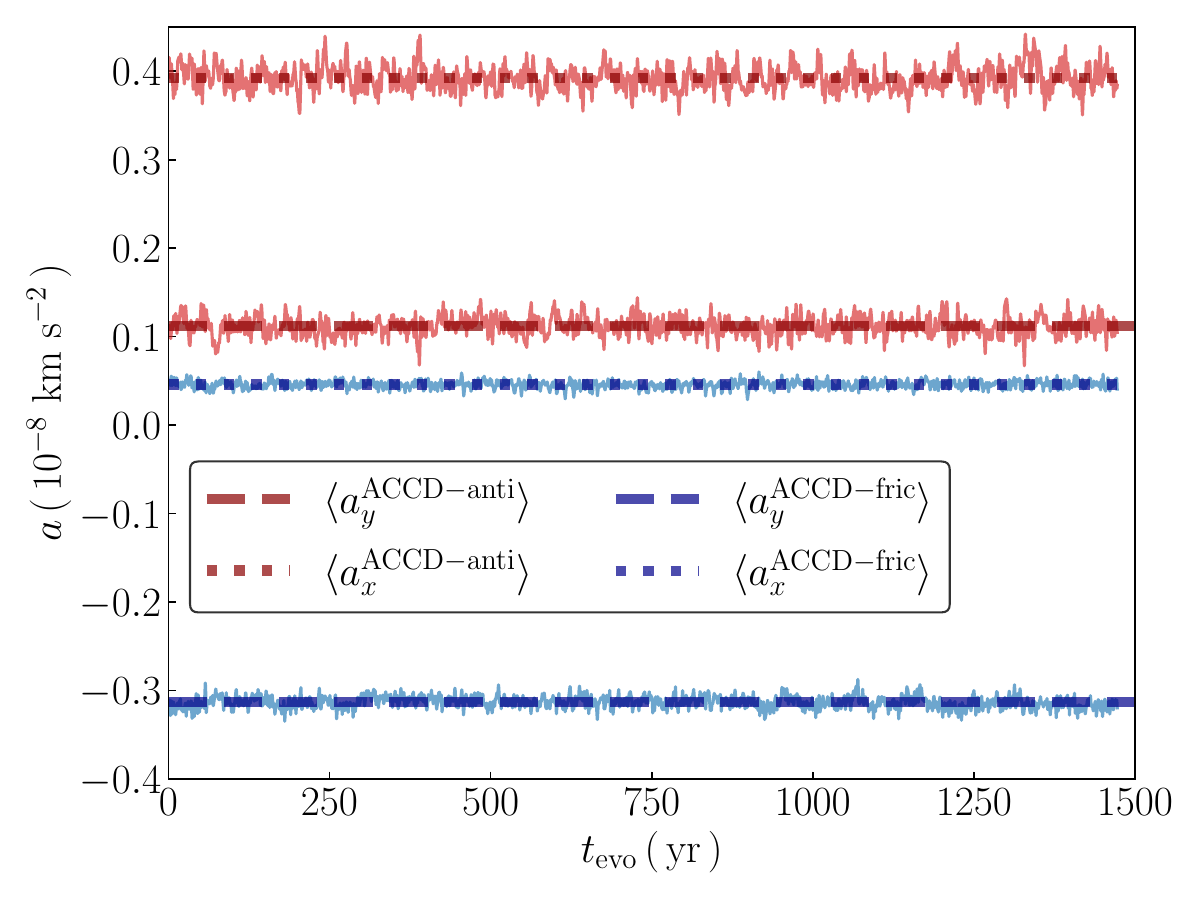}
\caption{Similar to Figure~\ref{fig:Basic-a}, but for the
  ACCD-anti (red lines) and ACCD-fric (blue lines)
  models. The dashed lines represent acceleration in the
  $y$-direction, while the dotted lines represent
  acceleration in the $x$-direction.  }
  \label{fig:ACCD-a-mdot}     
\end{figure}

\subsection{AGN Star in Accretion Disk}
\label{sec:accretion-disk} 

  AGN disks are likely accreting. Using a
  representative accretion velocity
  $v_{\rm acc}=10^6~{\rm cm}~{\rm s}^{-1}$, Model ACCD
  simulate an outflowing star embedded in an accreting AGN
  disk. Figure~\ref{fig:ACCD-rho} presents the density
  distribution in the ACCD model.  Because of the
  significant $v_x$ of disk gas towards the negative
  direction, the head-wind structure becomes more aligned
  toward the disk center compared to the \response{FID-anti} model.

  As Figure~\ref{fig:ACCD-a-mdot} shows, the star with
  outflows experiences acceleration in both the $x$- and
  $y$-directions due to anti-friction, resulting in rapid
  outward migration from the AGN disk. In disks with
  stronger accretion, the radial component of the
  anti-friction acceleration becomes increasingly
  significant, causing the AGN star's trajectory to align
  more closely with the radial direction. The orbital radius
  of the star expands at a rate of
  $\langle\dot{r}_{\rm cir}\rangle \approx 1.1~{\rm km}~{\rm s^{-1}}$,
  which is comparable to the FID-anti model.  In absence of
  stellar outflows, the star experiences dynamical friction
  from the ambient gas. The presence of accretion flow
  causes the star to experience acceleration in both the
  $-x$ and $-y$ directions. It is important to emphasize
  that the $y$-acceleration plays the dominant role in
  determining the orbital evolution of the star, while the
  influence of the $x$-acceleration is comparatively minor.

\subsection{The GAMMA Model with $\gamma = 4/3$}
\label{sec:gamma-section} 

\response{The fiducial model assumes that the star resides
  near the boundary between the two regions dominated by gas
  pressure and radiation pressure, respectively.  Given the
  fact that the outer disk is likely dominated by the
  radiation pressure, this sub-section simulates different
  cases adopting $\gamma = 4/3$ instead of $5/3$, while
  keeping all other parameters identical to the fiducial
  model. This case is hereafter referred to as the GAMMA
  model.}

\response{As shown in Figure~\ref{fig:Basic-a}, the
  GAMMA-anti model also undergoes acceleration due to
  anti-friction, although the magnitude is smaller than that
  in the FID-anti case. The comparison between
  Figures~\ref{fig:Basic-rho} and \ref{fig:Basic-gamma}
  indicates that the overdensity region leading the
  direction of motion is narrower in the GAMMA-anti model. A
  smaller $\gamma$ value renders the ambient gas more
  susceptible to the stellar outflow, particularly near the
  contact discontinuity. This effect partially disrupts the
  formation of an overdense region ahead of the star,
  weakening the anti-friction-induced acceleration. The
  fundamental mechanism of the interactions between the
  stellar outflow and ambient gas
  remains unchanged, and hence the results are mostly
  changed quantitatively rather than qualitatively. The
  quantitative impact of such change will be revisited in
  \S\ref{sec:black-hole}. }

\begin{figure} 
  \centering    
  \includegraphics[width=1.0\columnwidth]
  {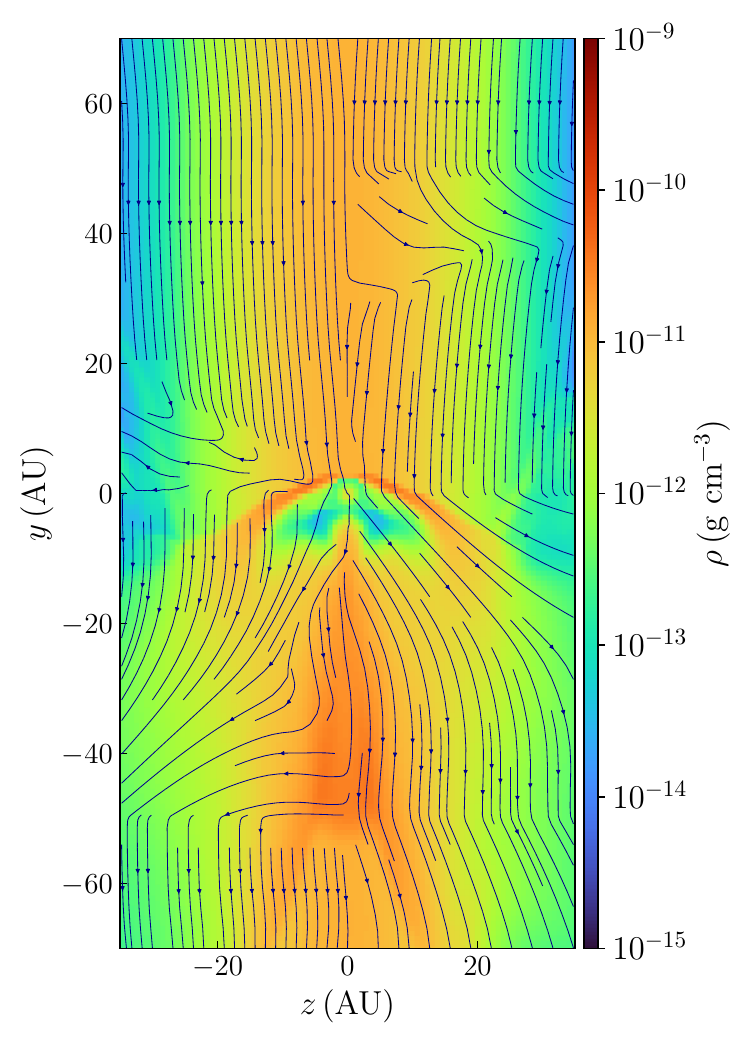}
  \caption{Gas density $\rho$ in the $x=0$ plane at
    $t_{\rm evo} = 0.1~P$ for the BH-jet-z models.}
  \label{fig:BH-jetz-rho}     
\end{figure}

\begin{figure} 
  \centering    
  \includegraphics[width=1.03\columnwidth]
{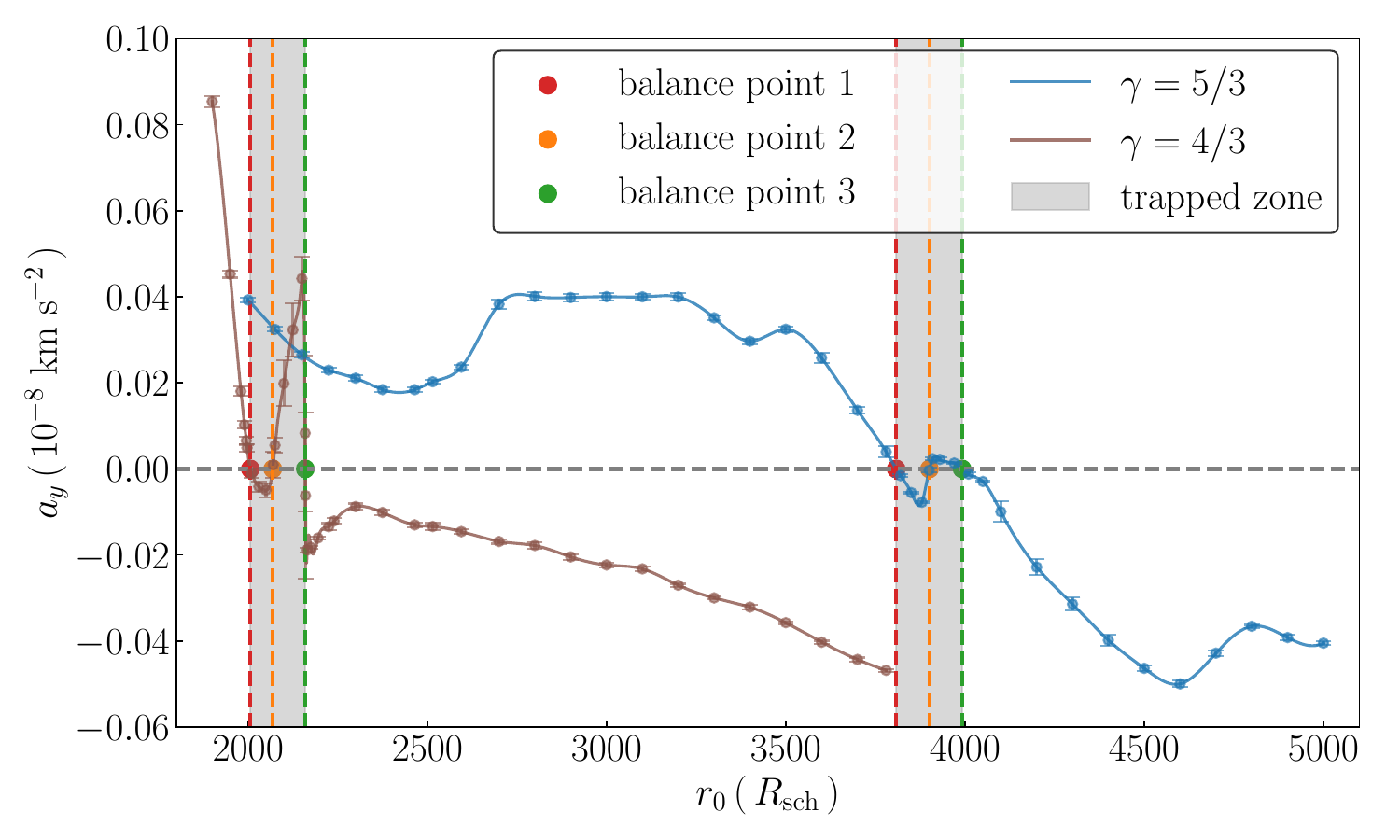}
\caption{Time-averaged azimuthal acceleration of the sBH in
  the BH-jet-z models for different values of $r_0$.}
  \label{fig:BH-a}     
\end{figure}

\section{Black Hole Orbit Dynamics in Accretion Disk}
\label{sec:black-hole}

  A type of extreme cases, which may be specifically
  interested by the studies on gravitational waves, involve
  sBHs with powerful outflows located at a distance of
  $3000~R_{\rm sch}$ from the center of a disk with
  significant accretion rate. Such simulation study, whose
  parameters are summarized in Table~\ref{tab:BH-property},
  have jet outflows aligned with the $z$-axis, and is
  referred to as BH-jet-z. The accretion disk is modeled by
  directly prescribing the velocity field of the disk gas.
  
  Figure~\ref{fig:BH-jetz-rho} presents a snapshot of the
  density distribution for the BH-jet-z model, whose impact
  of the sBH jet is distinct from the JET-z model.  An
  underdense region forms behind the sBH along its direction
  of motion, associated with the jet launched along the
  $z$-axis. Because the sBH resides near the disk center,
  the high inflow velocity in the $y$-direction suppresses
  the initially bidirectional jet launched along the
  $\pm z$-axis, confining the jet material to the trailing
  side of the sBH. Under these conditions, the requirements
  for anti-friction are satisfied, and the sBH is
  accelerated in the $y$-direction by anti-friction. The
  time-averaged accelerations is
  $\approx 0.40~{\rm km}~{\rm s^{-2}}$, and the
  corresponding orbital radius evolution rate is
  $\approx 1.83~{\rm km}~{\rm s^{-1}}$.


Based on the above analysis, an intuitive picture of the sBH
migration process can be proposed. If an sBH with jets
along the $z$-axis is embedded in the outer region of an
accretion disk, it would initially lose angular momentum due
to dynamical friction and migrate inward toward the AGN disk
center. As the sBH moves closer to the center, where the
ambient gas density and inflow velocity are sufficiently
high to confine the jet material to the trailing side of the
sBH, the anti-friction effect in the azimuthal direction
becomes effective. Consequently, the sBH gains angular
momentum and transitions to outward migration. Under this
interplay between inward and outward migration, the sBH is
expected to eventually settle at an equilibrium location
within the disk.

To explore the location of this trapped zone, a series of
simulations was carried out by varying only the distance
between the sBH and the disk center $r_0$, while keeping all
other parameters identical to BH-jet-z model. The value of
$r_0$ is varied in the range of $2000\,-\,5000~R_{\rm sch}$
to identify the radius at which $a_y = 0$. The results are
shown in Figure~\ref{fig:BH-a}. As $r_0$ increases, the
acceleration from anti-friction is gradually suppressed, and
beyond a certain distance, the anti-friction effect vanishes
entirely. Three balance points are showed in the figure: the
red and green points correspond to stable equilibrium, while
the yellow point represents an unstable one. Under the
current parameter setup, the sBH is found to be trapped
within the range of $3808\,-\,4006~R_{\rm
  sch}$. \response{The case with $\gamma = 4/3$ is also
  examined, and the results show overall trends consistent
  with those shown in Figure~\ref{fig:BH-a}, though the
  trapped zone shifts much closer to the disk center.} It is
worth noting that, as discussed in
\S\ref{sec:result-fiducial}, the sBH–gas interaction is also
influenced by factors such as the radial pressure gradient
and accretion velocity. Depending on the specific
conditions, the sBH may either gain or lose angular
momentum, resulting in outward migration or inward drift
toward the disk center. The location of the trapped zone is
also sensitive to parameter choices and may shift
accordingly. These scenarios will be explored in future
studies focusing on the parameters representing different
astrophysical systems with detailed measurements.


\begin{deluxetable}{lr}
  \tablecolumns{2}
  \tabletypesize{\footnotesize}
  \tablewidth{0pt}
  \setlength{\tabcolsep}{10pt}
  \tablecaption{Initial Properties of the BH-jet-z Models}
  \label{tab:BH-property}
  \tablehead{ \multicolumn{1}{l}{Item} &
    \multicolumn{1}{r}{Value} } \startdata
  \textbf{Disk Properties} & \\
  $M$ & $10^8~M_\odot$ \\
  $\rho_0$ & $\sim3.5\times10^{-11}~{\rm g~cm^{-3}}$ \\
  $c_{\rm iso}$ & \response{$10^6~{\rm cm~s^{-1}}$} \\
  $x_{\rm p}$ & $90~{\rm AU}$ \\
  $v_{\rm acc}^*$ & $-1\times10^6~{\rm cm~s^{-1}}$ \\
  \textbf{Stellar Properties} & \\
  $m$ & $10~M_\odot$ \\
  $r_0$ & $3000~R_{\rm sch}$ \\
  $\dot{m}$ & $3\times10^{-3}~M_\odot~{\rm yr^{-1}}$ \\
  $T_{\rm outflow}$ & $\sim5.7\times10^4~{\rm K}$ \\
  $v_{\rm src}$ & $1\times10^8~{\rm cm~s^{-1}}$ \\
  $r_{\rm soft}/r_{\rm src}$ & $\sim3$ \\
  Outflow Type$^\dagger$ & Jet \\
  \textbf{Simulation Parameters} & \\
  $L_x, L_y, L_z$ & $100, 200, 100~{\rm AU}$ \\
  Resolution $(N_x, N_y, N_z)$ & $128, 256, 128$ \\
  $t_{\rm evo}$ & $3~P$ \\
  \enddata \tablecomments{
    $^*$: The negative sign indicates that the disk is
    undergoing accretion. \\ 
    $^\dagger$:In the BH-jet-z model, the outflow is
    directed along the $\pm z$ axis as a jet.}
\end{deluxetable}

\section{Discussion and Summary}
\label{sec:summary}

This paper studies the dynamical evolution of stellar
objects, including stars and black holes with outflows,
located in the relatively outer region of AGN disks. In
general, dynamical friction causes angular momentum loss,
driving the AGN star toward the disk center. However, once
the outflows satisfy some conditions that enables
anti-friction, such stellar objects can gain angular
momentum from the ambient gas instead, leading to an outward
migration away from the AGN disk center.

\subsection{Impacts of the Anti-friction on AGN Star
  Dynamical Evolution}
\label{sec:summary-star}

Stars embedded within gaseous disks can excite spiral
density waves, and the resulting net torque from the ambient
gas, through dynamical friction, drives orbital
migration. Depending on the stellar mass and the presence or
absence of a gap in the ambient gas, migration has been
classically categorized into three types: Type I, Type II,
and Type III migration \citep{1979MNRAS.186..799L,
  1979MNRAS.188..191L, 1986ApJ...309..846L,
  1986Icar...67..164W, 2003ApJ...588..494M}. Although
dynamical friction has been extensively studied, the
phenomenon of anti-friction has received comparatively
little attention. It is also noted that, when supersonic
outflows clear up a region surrounding the embedded star,
the conclusions on the direction of orbital migration is
more robust by getting rid of the subtleties in the star
vincinities that could be dominant in ``traditional''
studies of migration.

In a stable anti-friction scenario, the high-density region
formed by the interaction between the stellar outflow and
the ambient gas consistently remains ahead of the star's
motion, while a low-density region persists behind it. This
configuration enables AGN stars with sufficiently strong
outflows to experience a more sustained and stable outward
migration compared to those undergoing conventional
migration mechanisms. The key factor in establishing
anti-friction lies in the balance between the ram pressure
of the stellar outflow and that of the ambient gas
\citep{2020MNRAS.494.2327L}. If the ambient gas density or
velocity is too low or if the stellar outflow is excessively
powerful, this balance cannot be achieved, and anti-friction
fails to operate effectively. Consequently, the strength of
anti-friction is closely tied to the physical properties of
the gas in the AGN disk, such as the radial pressure
gradient.

\response{Admittedly, there are several several caveats in
  this work related to the adopted stellar wind model, which
  should be addressed in future works. The stellar structure
  is simplified by modeling the star as a softened
  gravitational potential with artificially imposed
  outflows, without resolving realistic stellar
  envelopes. For the strong outflows at super-Eddington
  luminosities, two primary mechanisms are required, (1)
  additional energy release from the stellar interior, or
  (2) convective inefficiency near opacity bumps
  \citep{2017MNRAS.472.3749O,2012ASSL..384..275O}. The first
  mechanism is difficult to sustain over long timescales,
  while a necessary condition for the second is that the
  opacity bump region be optically thick, which requires
  very massive stars. Even under optimistic assumptions for
  isolated stars, continuum-driven mass loss rates exceeding
  $\sim 10^{-3}~M_\odot~{\rm yr}^{-1}$ are only possible
  when stellar masses exceed $40$-$60\,M_\odot$
  \citep{2024ApJ...974..270C}. Recent radiative hydrodynamic
  simulations indicate that for massive stars
  ($\sim 50\,M_\odot$) embedded in AGN-like environments,
  opacity-bump-induced mass loss is often suppressed by high
  accretion rates of $\sim 0.01~M_\odot~{\rm yr}^{-1}$
  \citep{2024ApJ...974..106C,2025ApJ...987..188C}. 
}

Accretion disks are vertically supported against the central
gravitational force primarily by radial pressure gradients,
which provide an upward force balancing gravitational
compression \citep{1973A&A....24..337S}. This pressure
support fundamentally determines the scale height and
overall structure of the disk. In AGN disks, the dominant
source of pressure varies with distance from the central
black hole: radiation pressure dominates the innermost
regions, whereas gas pressure and disk self-gravity become
more critical at larger radii
\citep{1986bhwd.book.....S,2003MNRAS.339..937G}. These
radial variations imply that the pressure gradient is not
uniform and can feature localized maxima or sharp
transitions. Recent studies, such as
\citet{2020ApJ...900...25J}, have demonstrated that the iron
opacity bump can increase the effective radiation pressure
for a given vertical flux, resulting in the formation of
pressure maxima within the disk. Variations in the radial
pressure profile have significant dynamical implications for
stars embedded in AGN disks. Specifically, the pressure
gradient modulates the relative inflow velocity of ambient
gas against the embedded star, which in turn influences the
efficiency of anti-friction. This study investigates the
role of the pressure gradient and finds that a weaker
pressure gradient reduces the relative velocity between the
star and the gas, thereby enhancing the anti-friction
acceleration. However, if the pressure gradient becomes too
small to maintain a stable head-wind structure around the
star, the anti-friction effect is substantially suppressed,
consequently hindering the outward migration of the AGN
star.

\response{ More realistic AGN disk models should also
  emphasize turbulence as a key role in shaping disk gas
  dynamics. Turbulence driven by magnetorotational
  instability is an important candidate mechanism of angular
  momentum transfer and overall disk dynamics in various
  types of accretion disks \citep{1998RvMP...70....1B,
    2004ApJ...602..595J,2011ARA&A..49..195A}, whose
  resulting gas density fluctuations could possibly disrupt
  the spatial configurations of overdense and underdense
  regions around the concerned moving objects, potentially
  enhancing or suppressing the dynamical friction effect,
  either positive or negative. Such disturbances may
  introduces stochasticity into the orbital evolution of
  embedded stars. As this current work does not account for
  turbulences, future works shall incorporate a turbulent
  background to reflect realistic AGN disk conditions more
  consistently.  }

\subsection{AGN sBH Migration}
\label{sec:summary-BH}

AGN disks have been recognized as promising environments for
the formation of extreme mass ratio inspirals (EMRIs)
\citep{2017ApJ...835..165B, 2020MNRAS.493.3732D,
  2021PhRvD.104f3007P, 2021PhRvD.103j3018P}. Stellar-mass
black holes (sBHs) can either originate from massive stars
formed at the disk outskirts or be captured from the
surrounding nuclear star cluster
\citep{1991MNRAS.250..505S,2004ApJ...608..108G}. Once
embedded within the disk, sBHs may migrate inward toward the
central SMBH due to various
hydrodynamic effects, including density wave interactions
\citep{1993ApJ...409..592A}, thermal torques
\citep{2024MNRAS.530.2114G}, and head- or tail-wind torques
in geometrically thick disks \citep{1993ApJ...411..610C,
  2011PhRvD..84b4032K, 2020ApJ...897..142S}. Those sBHs
reaching sufficiently close to the SMBH can become EMRIs,
and recent studies suggest that these wet EMRIs could
contribute significantly to the overall cosmic EMRI
population \citep{2021PhRvD.103j3018P, 2021PhRvD.104f3007P,
  2023MNRAS.521.4522D}. Hydrodynamic and three-body
simulations by \citet{2024arXiv241116070P} further
demonstrated that sBHs embedded in AGN disks can migrate in
tandem with gap-opening intermediate-mass black holes
(IMBHs), leading to sequential IMRI-IMRI formations or an
EMRI event followed by an IMRI, highlighting the intricate
interplay between migration and dynamical interactions.

This work reveals that, under the influence of
anti-friction, sBHs can extract angular momentum from the
ambient gas, initiating outward migration within the
accretion disk. In extreme cases where the outflow forms a
strong jet, the sBHs could experience significant torque by
azimuthal acceleration. If such an sBH initially migrates
inward from the outer regions of the disk, the interplay
between inward and outward migration could lead to its
eventual trapping near an equilibrium radius. This dynamic
process may result in the accumulation of multiple sBHs in
these regions, promoting binary formation and ultimately
leading to mergers.

\subsection{Future Works}
\label{sec:summary-future}

Due to the limitations in physical modeling, several caveats
remain in this study that need further investigation in
future works. The parameters adopted for all models were
chosen as fixed representative values, which may not fully
capture the diversity of the realistic scenarios. Future
work should incorporate observational constraints to refine
model setups. For the clarity of computation and analyses
procedcures, the approximation of adiabatic orbital
evolution is adopted, in which the stellar orbit and mass
are assumed to remain constant in the simulations, each of
which represents one single snapshot. The dynamical
evolution of stellar conditions, including mass, outflow
intensity, and orbital radius, would provide more realistic
results, especially when studying the anti-friction effects
in global simulations. 
In global cases, perturbations in the stellar wake could
traverse the entire orbital path and reinteract with the
star itself, potentially altering the migration
mechanism. Whether this feedback loop has a significant
impact on AGN star migration remains uncertain, which is
another interesting and important issue that worths further
investigations.

Binary systems embedded in disks are also of possible
interest, in which the physical scenario becomes
substantially more complex. Previous studies, such as the
series of works by \citet{2022MNRAS.517.1602L,
  2023MNRAS.522.1881L, 2024MNRAS.529..348L}, have
systematically examined the hydrodynamic evolution of
stellar-mass black hole binaries embedded in AGN disks,
analyzing how various physical parameters influence their
migration and eventual mergers. If both components of a
binary system produce outflows, interactions between the two
outflows, coupled with gas-star interactions, could
significantly impact the binary's orbital evolution. For
example, \citet{2022ApJ...932..108W} performed simulations
of AGB star–outflowing pulsar systems and found that a dense
and slow outflow can exert a positive torque on the
binary. This torque leads to orbital expansion by more than
$10\%$.  It is therefore expected that binary evolution
within AGN disks would also be strongly influenced by
outflow–outflow interactions, which will be investigated in
future work.

In addition, the types of astrophysical systems could extend
far beyond this study, which primarily focuses on the impact
of anti-friction on the AGN stars migration. Similar
physical mechanisms may exist in other astrophysical
scenarios. For instance, \citet{2025ApJ...978...87L}
investigated the role of anti-friction in the migration of
stars within open clusters. In environments such as
protoplanetary disks, where the gas density is significantly
lower than in AGN disks, the potential impact of
anti-friction on planetary migration remains non-negligible
if the planets are capable of generating sufficiently strong
outflows. Future studies will extend the investigation of
anti-friction to such systems.

\begin{acknowledgments}
  M. Liu and L. Wang appreciates the computational resources
  provided by the Kavli Institute of Astronomy and
  Astrophysics at Peking University.  We also thank our
  colleagues Ruobin Dong, Xian Chen, D. N. C. Lin, Xinyu Li and Erlin Qiao for helpful discussions and suggestions.
\end{acknowledgments}

\bibliographystyle{aasjournal}
\bibliography{NDF-AGNstar}{}

\end{document}